\documentclass[a4paper,USenglish,cleveref, thm-restate]{lipics-template/lipics-v2021}

\usepackage{cite}
\usepackage{amsmath,amssymb,amsfonts}
\usepackage{algorithmic}
\usepackage{graphicx}
\usepackage{textcomp}
\usepackage[table]{xcolor}
\usepackage{siunitx}
\usepackage{array}

\usepackage{amsthm}
\usepackage{etoolbox}
\usepackage{thmtools}
\usepackage{xspace}
\usepackage{dsfont}
\usepackage[capitalise]{cleveref}
\let\url\texttt

\usepackage[linesnumbered,ruled]{algorithm2e}

\newenvironment{myalgorithm}[1][!t]{
	\let\oldnl\nl
	\newcommand{\nonl}{\renewcommand{\nl}{\let\nl\oldnl}}
	
	\SetCommentSty{myalgorithmCommSty}
	\SetDataSty{myKwDateSty}
	\DontPrintSemicolon
	\SetAlgoLined\SetAlgoNoEnd
	\SetArgSty{normalfont}
	\SetKwInOut{Input}{Input}
	\SetKwInOut{Output}{Output}
	\SetKw{Continue}{continue}
	\SetKwFor{ParForEach}{for each}{in parallel}{}
	\SetKwFor{ParFor}{for}{in parallel}{}
	\SetKwFor{DoWhile}{do}{}{while}
	\SetKwRepeat{Do}{do}{while}
	\SetKw{Sequentially}{sequentially}
\begin{algorithm}[#1]\footnotesize
}{\end{algorithm}}

\usepackage{tikz}
\usetikzlibrary{calc}

\newtheoremstyle{mythmstyle}
{.75em} 		
{.75em} 		
{\normalfont}	
{} 				
{\bfseries} 	
{.} 			
{.75em} 		
{} 				
\theoremstyle{mythmstyle}

\declaretheorem[style=mythmstyle,name=Definition]{mydef}
\declaretheorem[style=mythmstyle,name=Theorem]{mythm}
\declaretheorem[style=mythmstyle,name=Lemma]{mylem}
\declaretheorem[style=mythmstyle,name=Observation]{myobs}
\declaretheorem[style=mythmstyle,name=Corollary]{mycol}

\def\diagonal#1{\rotatebox{90}{#1}} 

\def\es{\ensuremath{\sigma}}
\def\1{\mathds{1}}
\def\degseq{\ensuremath{\mathbf{d}}}
\def\alldegseq{\ensuremath{\mathcal G(\degseq)}}

\def\ie{i.e.,\xspace}
\def\eg{e.g.\xspace}
\def\etal{~et\,al.\xspace}
\def\prob#1{\ensuremath{\mathbb P\left[#1\right]}}
\def\Oh#1{\ensuremath{\mathcal O \left(#1\right)}}
\def\nproc{\ensuremath{\mathcal P}}

\def\algoname#1{\textsc{#1}\xspace}
\def\esmc{\algoname{ES-MC}}
\def\gesmc{\algoname{G-ES-MC}}
\def\fastes{\algoname{SeqES}}
\def\globes{\algoname{SeqGlobalES}}
\def\pares{\algoname{NaiveParES}}
\def\trueparbatch{\algoname{ParallelSuperstep}}
\def\truepares{\algoname{ParES}}
\def\trueparglobes{\algoname{ParGlobalES}}

\def\figscale{0.8}
\newenvironment{condRotTab}{}{}

\hideLIPIcs

\title{Parallel Global Edge Switching for the Uniform Sampling of Simple Graphs with Prescribed Degrees\footnote{Accepted manuscript. Link to final version: \url{https://doi.org/10.1016/j.jpdc.2022.12.010}}}

\author{Daniel Allendorf}{Goethe University Frankfurt, Germany}{daniel@ae.cs.uni-frankfurt.de}{https://orcid.org/0000-0002-0549-7576}{}
\author{Ulrich Meyer}{Goethe University Frankfurt, Germany}{umeyer@ae.cs.uni-frankfurt.de}{}{}
\author{Manuel Penschuck}{Goethe University Frankfurt, Germany}{mpenschuck@ae.cs.uni-frankfurt.de}{https://orcid.org/0000-0003-2630-7548}{}
\author{Hung Tran}{Goethe University Frankfurt, Germany}{htran@ae.cs.uni-frankfurt.de}{}{}

\titlerunning{Parallel Global Edge Switching for the Uniform Sampling of Graphs}

\authorrunning{D.~Allendorf, U.~Meyer, M.~Penschuck and H.~Tran}

\Copyright{Daniel Allendorf}



\supplement{Source code: \url{https://github.com/manpen/global-edge-switching}.}

\begin{document}

\maketitle

\begin{abstract}%
	The uniform sampling of simple graphs matching a prescribed degree sequence is an important tool in network science, e.g. to construct graph generators or null-models.
Here, the Edge Switching Markov Chain (ES-MC) is a common choice.
Given an arbitrary simple graph with the required degree sequence, ES-MC carries out a large number of small changes, called edge switches, to eventually obtain a uniform sample.
In practice, reasonably short runs efficiently yield approximate uniform samples.

In this work, we study the problem of executing edge switches in parallel.
We discuss parallelizations of ES-MC, but find that this approach suffers from complex dependencies between edge switches.
For this reason, we propose the Global Edge Switching Markov Chain (G-ES-MC), an ES-MC variant with simpler dependencies.
We show that G-ES-MC converges to the uniform distribution and design shared-memory parallel algorithms for ES-MC and G-ES-MC.
In an empirical evaluation, we provide evidence that G-ES-MC requires not more switches than ES-MC (and often fewer), and demonstrate the efficiency and scalability of our parallel G-ES-MC implementation.
\end{abstract}

\section{Introduction}
In network science there are various measures, so-called centralities, to quantify the importance of nodes~\cite{DBLP:journals/im/BoldiV14}.
The degree centrality, for instance, suggests that a node's importance is proportional to its degree, \ie the number of neighbors it has (see also \cite{barabasi2016network}).
This leads to the natural question whether graphs with matching degrees share structural properties.
While this is not the case in general, a reoccurring task in practice is to quantify the statistical significance of some property observed in a network.
Given an observed graph with degrees~$\degseq$, a popular null-model is the uniform distribution over all simple graphs $\alldegseq$ with matching degrees~\cite{Cobb2003,Itzkovitz2003,DBLP:conf/asunam/SchlauchZTA15}.

In this context, the \emph{Edge Switching Markov Chain} (\esmc) is a common choice to obtain an approximate uniform sample from $\alldegseq$.
In each so-called \emph{edge switch}, two edges are selected uniformly at random and modified by exchanging their endpoints.
This process preserves the degrees of all nodes involved.
We further keep the graph simple by rejecting all edge switches that introduce self-loops or multi-edges.

There exist different variants of \esmc catering to various graph classes (\eg Carstens~\cite{CarstensPhd} considers directed/undirected graphs, with/without loops, with/without multi-edges).
Here, we focus on simple and undirected graphs.
It is, however, straight-forward to adopt our findings to the other cases (some of which even lead to easier algorithms).

\subsection{Related Work}
Various methods to obtain a graph from a prescribed degree sequence have been studied~\cite{DBLP:journals/corr/abs-2003-00736}.

Havel~\cite{Havel1955} and Hakimi~\cite{doi:10.1137/0110037} independently lay the foundation for a deterministic linear time generator.
The algorithm, however, does not yield random graphs; while randomizations~(\eg \cite{DBLP:journals/im/BlitzsteinD11,DBLP:conf/bigdataconf/BhuiyanKM17}) are available, they produce non-uniform samples.

The Chung-Lu Model~\cite{chung2002connected} constructs graphs that match the prescribed degrees only in expectation.
Under reasonable assumptions~\cite{DBLP:journals/corr/abs-2003-00736} it can be sampled in linear time~\cite{DBLP:journals/datamine/MorenoPN18}.

The Configuration Model~\cite{DBLP:journals/jct/BenderC78} outputs a random, but possibly non-simple, graph in linear time;
adding rejection-sampling yields simple graphs in polynomial time if the maximum degree is $\Oh{\sqrt{\log n}}$ (cf. \cite{bekessy1972asymptotic, DBLP:books/ox/Newman10,bollobas1985random, DBLP:journals/ejc/Bollobas80}).
Efficient and exact uniform generators can be obtained by adding a repair step between the Configuration Model and the rejection-sampling to boost the acceptance probability.
Such algorithms are available for several degree sequences classes including bounded regular graphs or power-law sequences with sufficiently large exponents~\cite{DBLP:journals/jal/McKayW90,DBLP:conf/focs/GaoW15,DBLP:conf/focs/ArmanGW19}.

Further a plethora of Markov Chain Monte-Carlo (MCMC) algorithms have been proposed and analyzed (e.g.~\cite{DBLP:journals/tcs/JerrumS90,DBLP:journals/cpc/CooperDG07, DBLP:conf/alenex/GkantsidisMMZ03, DBLP:conf/soda/Greenhill15, DBLP:journals/rsa/KannanTV99, DBLP:conf/sigcomm/MahadevanKFV06, DBLP:conf/alenex/StantonP11, strona2014fast, verhelst2008efficient, DBLP:journals/compnet/VigerL16}).
In comparison to the aforementioned exactly uniform generators, these algorithms allow for larger families of degree sequences, topological restrictions (\eg connected graphs~\cite{DBLP:conf/alenex/GkantsidisMMZ03, DBLP:journals/compnet/VigerL16}), or more general characterizations (\eg joint degrees~\cite{DBLP:conf/alenex/StantonP11, DBLP:conf/sigcomm/MahadevanKFV06}). 
Switch Markov Chains such as \esmc have been shown to be rapidly mixing for classes of undirected graphs such as bounded-degree or power-law graphs (c.f.~\cite{DBLP:journals/cpc/CooperDG07, 10.1371/journal.pone.0131300, DBLP:journals/tcs/GreenhillS18, ERDOS2022103421, DBLP:journals/rsa/AmanatidisK20, DBLP:journals/dam/GaoG21, DBLP:journals/ejc/ErdosGMMSS22}).
While these bounds remain impractical for everyday use due to the high degrees of the polynomials involved, empirical studies suggest that a number of steps linear in the number of edges yields samples which are sufficiently uncorrelated to the input graph \cite{Milo2003,DBLP:conf/alenex/GkantsidisMMZ03,DBLP:conf/waw/RayPS12}.

The only prior parallelization of \esmc we are aware of was given by \cite{DBLP:journals/jpdc/BhuiyanKCM17}.
While the proposed algorithm has the advantage that it can be used in a distributed setting, it only avoids conflicts between edge switches that arise due to concurrent accesses to the same edge.
This however is not enough to ensure that the algorithm faithfully implements \esmc, e.g. the graphs generated by the parallel process differ from the ones generated by a sequential process.
To address this issue the authors conduct an empirical error analysis.
However, the issue remains that the graphs generated by such a process may not converge to the uniform distribution due to asymmetrical transition probabilities \cite{9820710}.
In addition, \cite{DBLP:conf/esa/CarstensH0PTW18} proposed an external memory parallelization of the Curveball Markov Chain for sampling undirected graphs.
Note however, that this algorithm requires sorting as a subroutine, which may be too expensive for usage in internal memory.
In contrast to switch Markov Chains, Curveball has also not been shown to be rapidly mixing for undirected graphs, and despite preliminary empirical results \cite{DBLP:conf/esa/CarstensH0PTW18}, it remains an open question to relate the mixing times of Curveball and \esmc for undirected graphs.



\subsection{Our Contribution}

We propose the Global Edge Switching Markov Chain (\gesmc), a new switch Markov Chain for sampling simple undirected graphs with prescribed degrees and show that it converges to the uniform distribution.
This Markov Chain is designed with parallel algorithms in mind and exhibits less complex dependencies between edge switches.
Consequently, we describe and analyze an exact shared-memory parallel algorithm for \gesmc.
In addition, we describe an exact parallel algorithm for \esmc.

In an empirical study, we provide evidence that \gesmc mixes faster than or equally well as standard \esmc, and demonstrate the efficiency and scalability of our parallel \gesmc implementation.
To the best of our knowledge, our implementation outperforms all openly available solutions by up to two orders of magnitude using 32 threads.

\subsection{Outline}

The article is structured as follows.
\cref{sec:prelim} covers the notation used and the necessary background for our results, such as a definition of the Edge Switching Markov Chain (\esmc).
In \cref{sec:g-es-mc}, we introduce the Global Edge Switching Markov Chain (\gesmc) and prove its convergence to the uniform distribution.
\cref{sec:steady-algorithm} contains a description of our parallel algorithms for \esmc and \gesmc, and a formal analysis of the \gesmc algorithm.
We discuss implementation details of the parallel \gesmc algorithm, as well as sequential and parallel baseline/reference implementations, in \cref{sec:implementation}.
In \cref{sec:experiments}, we relate the mixing time of \esmc and \gesmc empirically and investigate the efficiency and scalability of our implementations.
\cref{sec:conclusions} concludes the article with a summary of the results and an outlook on potential future work.

\section{Preliminaries}
\label{sec:prelim}
\subsection{Notation and Definitions}
Define the short-hands $[k..n] := \{k, \ldots, n\}$ and $[n] := [1..n]$.
A graph $G = (V, E)$ has $n$ nodes $V=\{v_1, \ldots, v_n\}$ and $m$ undirected edges $E$.
We assume that edges are indexed (\eg by their position in an edge list) and denote the $i$-th edge of a graph as $E[i]$.
Given an undirected edge $e = \{v_i, v_j\}$ we denote a directed representation as $\vec e$.
We treat both as the same object and defining one implies the other; 
we default to the canonical orientation $\vec e = (v_{\min(i, j)}, v_{\max(i, j)})$ whenever the direction is ambiguous.
An edge $(v, v)$ is called a \emph{loop} at $v$; an edge that appears more than once is called a \emph{multi-edge}.
A graph is \emph{simple} if it contains neither multi-edges nor loops.

Given a graph $G=(V,E)$ and a node $v \in V$, define the \emph{degree} $\deg(v) = \left|\{ u\, \colon \{u,v\} \in E \}\right|$ as the number of edges incident to node $v$.
Let $\degseq = (d_1, \dots, d_n) \in \mathbb N ^ n$ be a \emph{degree sequence} and denote $\alldegseq$ as the set of simple graphs on $n$~nodes with $\deg(v_i) = d_i$ for all $v_i \in V$.
The degree sequence $\degseq$ is \emph{graphical} if $\alldegseq$ is non-empty.

A commonly considered class of graphs are power-law graphs where the degrees follow a power-law distribution.
To this end, let $\textsc{Pld}([a..b], \gamma)$ refer to an integer \underline{P}ower-\underline{l}aw \underline{D}istribution with exponent ${-}\gamma \in \mathbb R$ for $\gamma \ge 1$ and values from the interval $[a..b]$; let $X$ be an integer random variable drawn from $\textsc{Pld}([a..b], \gamma)$, then $\prob{X = k} \propto k^{-\gamma}$ (proportional to) if $a \le k \le b$ and $\prob{X = k} = 0$ otherwise.

\subsection{The Edge Switching Markov Chain (\esmc)}
\begin{figure}
	\begin{center}
		\begin{tikzpicture}[
		node/.style={draw, minimum height=1.2em, circle, inner sep=0, font=\footnotesize},
		edge/.style={draw, ->},
		emph edge/.style={edge, thick, red},
		label/.style={anchor=north, align=center, font=\footnotesize}
	]
	\def\dx{3em}
	\foreach \i in {0, 1, 2, 3} {
		\node[node] (a\i) at (6.8em*\i, 0) {A};
		\node[node, xshift=\dx] (b\i) at (a\i) {B};
		\node[node, yshift=-\dx] (x\i) at (a\i) {X};
		\node[node, yshift=-\dx, xshift=\dx] (y\i) at (a\i) {Y};
		\coordinate (l\i) at ($(a\i) + (0.5*\dx, -1.3*\dx)$);
	}

	\foreach \i in {0, 1, 2} {
		\path[draw, dotted] (6.8em*\i + 4.9em, 0em) to ++(0, -3em);
	}

	\path[edge] (a0) to node[above]{$e_1$} (b0);
	\path[edge] (a0) to node[right]{$e_5$} (x0);
	\path[edge] (x0) to node[below]{$e_2$} (y0);
	\node[label] at (l0) {Input $G$};
	
	\path[emph edge, bend right=15]  (a1) to node[left]{$e_3$} (x1);
	\path[emph edge, bend left=15, opacity=0.5] (a1) to node[right]{$e_5$} (x1);
	\path[edge] (b1) to node[right]{$e_4$} (y1);
	\node[label, color=red] at (l1) {
		$\sigma\big((e_1, e_2), 0\big)$ \\
		rejected: multi-edge
	};

	\path[edge] (a2) to node[above]{$e_3$} (y2);
	\path[edge, opacity=0.5] (a2) to node[left]{$e_5$} (x2);
	\path[edge] (b2) to node[below]{$e_4$} (x2);
	\node[label, color=green!50!black] at (l2) {
		$\sigma\big((e_1, e_2), 1\big)$ \\
		accepted
	};

	\path[emph edge, out=20, in=340, looseness=10]  (a3) to node[above]{$e_3$} (a3);
	\path[edge] (b3) to node[right]{$e_4$} (x3);
	\path[edge, opacity=0.5] (x3) to node[below]{$e_2$} (y3);
	\node[label, color=red] at (l3) {
		$\sigma\big((e_1, e_5), 0\big)$ \\
		rejected: loop
	};
	
\end{tikzpicture}
	\end{center}
	\vspace{-1em}
	\caption{
		Edge Switch on an undirected graph $G = (V, E)$.
		To avoid ambiguity, we indicate for each $e \in E$ the orientation $\vec e$ used in \cref{def:ses}.
	}
	\label{fig:single-es}
\end{figure}
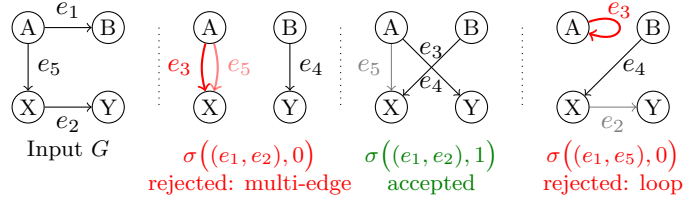

\begin{mydef}[Edge Switch, \esmc]\label{def:ses}
	Let $G=(V,E) \in \alldegseq$ be a simple undirected graph.
	We represent an edge switch~${\sigma = (i, j, g)}$ by two indices $i, j \in [m]$ and a direction bit~$g$.
	Then, we compute $G' \in \alldegseq$ based on $\sigma$ as follows:
	\begin{enumerate}
		\item Let $e_1 = E[i]$ and $e_2 = E[j]$.
		\item Compute new edges $(\vec e_3, \vec e_4) = \tau(\vec e_1, \vec e_2, g)$ where
		\begin{equation*}
			\tau\big((u, v), (x, y), g\big) =
			\begin{cases}
				\big((u, x), (v, y)\big) & \text{if } $g = 0$ \\
				\big((u, y), (v, x)\big) & \text{if } $g = 1$
			\end{cases}	 .
		\end{equation*}
		\item \emph{Reject} if either of $e_3$ or $e_4$ is a loop or already exists in $E$;
		otherwise \emph{accept} and set $E[i] \gets e_3$ and $E[j] \gets e_4$.
	\end{enumerate}
	
	\noindent The \emph{Edge Switching Markov Chain} (\esmc) transitions from state $G$ to state $G' \in \alldegseq$ by sampling $i$, $j$, and $g$ uniformly at random. \hfill$\triangle$
\end{mydef}

\begin{myobs}\label{obs:ses-inverse}
	Any edge switch~$\es = (i, j, g)$ that translates $G$ into $G'$ can be reversed in a single step, \ie there exists an inverse edge switch $\tilde \sigma$ that translates $G'$ back into $G$.
	If $\es$ is rejected, it does not alter the graph ($G = G'$).
	Thus the claim trivially holds with $\tilde \sigma = \sigma$.
	For an accepted edge switch $\sigma$, it is easy to verify that $\tilde \sigma = (i, j, 0)$ reverses the effects of $\sigma$.
	Further observe that the probability of choosing $\sigma$ in state $G$ equals the probability of choosing $\tilde \sigma$ in state $G'$ \cite{DBLP:journals/dam/GaoG21}. \hfill$\triangle$
\end{myobs}

\section{Global Edge Switching Markov Chain (\gesmc)}
\label{sec:g-es-mc}
Hamann\etal~\cite{DBLP:journals/jea/HamannMPTW18} consider the out-of-order execution of a batch consisting of $\ell$~edge switches.
To this end, they classify the dependencies within the batch that arise if the switches were executed in-order.
We adopt this characterization distinguishing between source- and target dependencies:

\begin{mydef}[Source/target dependencies]
	Two switches $\sigma_1 = (i_1, j_1, \cdot)$ and $\sigma_2 = (i_2, j_2, \cdot)$ are \emph{source dependent} if they share at least one source index, \ie $\{i_1, j_1\} \cap \{i_2, j_2\} \ne \emptyset$.
	Two switches $(e_1, f_1) \gets \sigma(\cdot, \cdot, \cdot)$ and $(e_2, f_2) \gets \sigma(\cdot, \cdot, \cdot)$ have a \emph{target dependency} if they try to produce at least one common edge, \ie 
	$\{e_1, f_1\} \cap \{e_2, f_2\} \ne \emptyset$; this dependency is counted even if one or both edge switches are rejected.
	\hfill$\triangle$
\end{mydef}

Source dependencies can be modelled by a balls-into-bins process where edges correspond to bins and each edge switch throws a linked pair of balls into two bins chosen uniformly at random.
A source dependency arises whenever a ball falls into a non-empty bin.
Czumaj and Lingas~\cite{DBLP:journals/corr/abs-2108-11613} analyse this process in a different context.
Interpreted for \esmc, they show that for $\ell = m$ the longest source dependency chain has an expected length of $\Theta(\log m / \log\log m)$.
The distribution of target dependencies, on the other hand, depends on the graph's degree sequence as the probability that a random edge switch produces the edge $\{u, v\}$ is proportional to $\deg(u) \cdot \deg(v)$.

From an algorithmic point of view, source dependencies are more difficult to deal with if we want to process edge switches out-of-order (\eg for parallel execution).
Since each previous edge switch may or may not change the edge associated with the colliding edge index, the number of possible assignments may grow exponentially in the length of the dependency chain (if multiple chains cross).
Consequently, we need to either serialize such edge switches or accommodate all possible assignments.
In contrast, target dependencies only imply a binary predicate, namely whether a previous edge switch already introduced the target edge.

The above discussion suggests that it is most reasonable to parallelize \esmc by parallelizing batches of edge switches without source dependencies.
Naturally, the scalability of such a parallelization depends on the size of these batches.
For \esmc, each switch selects its source edges uniformly at random, and the probability of a source dependency between two edge switches is $\Theta(1 / m)$.
Thus, the expected batch size for \esmc is $\Theta(\sqrt{m})$.

As the expected batch size for \esmc is rather small, it is natural to ask if there exist other switch Markov Chains where the size is larger, and which thereby exhibit more parallelism.
To answer this question, we define a \emph{global switch} \footnote{We adapt the term \emph{global} from \cite{DBLP:journals/corr/CarstensBS16} where it is used to describe a variant of the Curveball Markov Chain.} --- a batch of up to $\lfloor m/2 \rfloor$ edge switches where each edge participates in an edge switch exactly once; conceptually, we place all edges into an urn and iteratively draw without replacement pairs of edges until the urn is empty; each pair implies an edge switch.
It is folklore to encode such a process in a permutation that captures the order of edges drawn~\cite{DBLP:journals/corr/abs-2003-00736}.

Similarly to techniques of \cite{DBLP:journals/corr/CarstensBS16,DBLP:conf/esa/CarstensH0PTW18}, our proof of \cref{thm:g-es-mixes} requires a small positive probability that any global switch collapses into a single switch.
We implement this by independently rejecting each switch with probability $P_L$.

\begin{mydef}[Global Switch, \gesmc]\label{def:ges}
	Let $G = (V, E)$ be a simple undirected graph. 
	A \emph{global switch} is represented by $\Gamma = (\pi, \ell)$ where $\pi$ is a permutation of $[m]$ and $\ell$ an integer with $0\le \ell \le \lfloor m/2\rfloor$.
	The global switch~$\Gamma$ consists of $\ell$ edge switches $\sigma_1, \ldots, \sigma_\ell$ that are executed in sequence, where $\sigma_k = (\pi(2k - 1),\, \pi(2k),\, g_k)$ and\ $g_k = \1_{\pi(2k - 1) < \pi(2k)}$, and where $\1$ denotes the indicator function.

	The \emph{Global Edge Switching Markov Chain} (\gesmc) transitions from graph~$G$ using a random global switch $\Gamma = (\pi, \ell)$.
	To this end, $\pi$ is drawn uniformly from all permutations on $[m]$ and $\ell$ is drawn from a binomial distribution of $\lfloor m/2 \rfloor$ trails with success probability $0 < P_L < 1$.\hfill$\triangle$
\end{mydef}

By selecting $\ell$ from a binomial distribution and executing the first $\ell$ edge switches of a random permutation, we simulate $\lfloor m/2 \rfloor$ edge switches that are each executed only with probability $1 - P_L$.
Also note, that the direction bits $g_k = \1_{\pi(2k - 1) < \pi(2k)}$ are independent and unbiased random bits because the permutation~$\pi$ is drawn uniformly at random.

\begin{mythm}\label{thm:g-es-mixes}
	Let $G \in \alldegseq$ be a simple undirected graph with degree sequence~$\degseq$.
	The Global Edge Switching Markov Chain started at $G$ converges to the uniform distribution on $\alldegseq$.
\end{mythm}

\begin{proof}
	Any Markov Chain that is irreducible, aperiodic and symmetric converges to a uniform distribution \cite[Th.~7.10]{DBLP:books/daglib/0012859}.
	We show that \gesmc has these three properties.
	
	For \textbf{irreducibility} we observe that whenever there exists an edge switch from state $A$ to $B$, there also exists a global switch from $A$ to $B$ (\eg if $\ell = 1$).
	The state graph of \esmc is therefore a subgraph of the state graph of \gesmc.
	In addition, both Markov Chains share the same set of states, \ie the set of all simple graphs with the given degree sequence.
	Then, since the state graph of \esmc is already strongly connected~\cite{rao1996}, so is the state graph of \gesmc, and thus both Markov Chains are irreducible.
	
	For \textbf{aperiodicity} we note that a global switch~$\Gamma = (\pi, \ell)$ may not alter the graph (\eg, if $\ell = 0$).
	Thus each state in the Markov Chain has a self-loop with strictly positive probability mass.
	This guarantees aperiodicity.

	It remains to show the \textbf{symmetry} of transition probabilities.
	Let $\mathcal S_{AB}$ be the set of global switches~$\Gamma$ that transform graph~$A$ into graph~$B$.
	Then the transition probability~$P_{AB}$ from $A$ to $B$ equals the probability of drawing a global switch from $\mathcal S_{AB}$, 
	\begin{equation*}
		P_{AB} = \sum_{\Gamma \in \mathcal S_{AB}} P(\Gamma),
		\vspace{-0.0em}
	\end{equation*}
	where $P(\Gamma)$ is the probability of selecting $\Gamma$.
	A global switch~$\Gamma = (\pi, \ell)$ is selected by drawing its permutation $\pi$ and executing $\ell$ edge switches.
	In particular, we have
	\begin{equation*}
		P(\Gamma) = \underbrace{\frac{1}{m !}}_\text{uniform $\pi$} \quad \cdot \quad \underbrace{\binom{\lfloor m/2\rfloor}{\ell} (1 - P_L)^\ell P_L^{\lfloor m/2\rfloor - \ell}}_\text{binomially distributed $\ell$}.
	\end{equation*}
	Observe that $P(\Gamma = (\pi, \ell))$ depends on $\ell$, but neither on a specific choice of $\pi$ nor the states $A$ and $B$.
	Thus the symmetry $P_{AB} = P_{BA}$ follows by establishing a bijection $\mu_{AB}$ between any forward global switch $\Gamma {=} (\pi, \ell) \in \mathcal S_{AB}$ and an inverse global switch $\tilde\Gamma {=} (\tilde \pi, \ell) \in \mathcal S_{BA}$ with matching $\ell$.
	
	We construct the bijection $\mu_{AB}$ as follows.
	For a global switch $\Gamma = (\pi, \ell)$ that executes the edge switches $\sigma_1, \dots, \sigma_\ell$ with $\sigma_k = (\pi(2k - 1),\ \pi(2k),\ g_k)$ in sequence, define the inverse global switch $\tilde{\Gamma} = (\tilde{\pi}, \ell)$.
	The global switch $\tilde{\Gamma}$ executes the inverse edge switches in reverse order, \ie $\tilde{\sigma}_{\ell - k + 1}$ recovers the effect of the forward edge switch $\sigma_{k}$.

	Recall that (i) the inverse of an accepted edge switch is given by a direction flag $g = 0$, and that (ii) the forward direction bit is defined as $g_k = \1_{\pi(2k - 1) < \pi(2k)}$.
	Thus, if $\sigma_k$ is legal and $g_k = 1$, we need to switch the order of the edge indices in the inverse switch $\tilde{\sigma}_{\ell - k + 1}$. This implies $\tilde \pi$ on positions $[2\ell]$. In particular, we have for $k \in [\ell]$:
	\begin{align*}
		&\Big(\tilde{\pi}(2[\ell {-} k {+} 1] - 1),\quad \tilde{\pi}(2 [\ell -  k + 1])\Big) = \\
		&\quad \begin{cases}
			\big( \pi(2k - 1), \quad \pi(2k) \big) & \text{$\sigma_k$ is illegal or $g_k = 0$} \\
			\big( \pi(2k), \quad \pi(2k - 1) \big) & \text{$\sigma_k$ is legal and $g_k = 1$} \\
		\end{cases}
	\end{align*}
	For the unused entries $i \in [2\ell{+1}..m]$ choose $\tilde \pi(i) = \pi(i)$.
\end{proof}

\subsection{Mixing Time of \gesmc in relation to \esmc}
\label{sec:mixing-times}
For simple graphs, the mixing time of variants of \esmc have been studied for many families of degree sequences (c.f.~\cite{DBLP:journals/cpc/CooperDG07, 10.1371/journal.pone.0131300, DBLP:journals/tcs/GreenhillS18, ERDOS2022103421, DBLP:journals/rsa/AmanatidisK20, DBLP:journals/dam/GaoG21}).
Recently, Erd{\"{o}}s et al.~\cite{DBLP:journals/ejc/ErdosGMMSS22} provide a survey that unifies many proofs on rapidly mixing Switching Markov chains on different types of degree sequences. 
Furthermore, for bipartite degree sequences a theoretical comparison between a variant of \esmc and the more recent Curveball Markov Chain~\cite{verhelst2008efficient} has been established~\cite{DBLP:conf/approx/CarstensK18}.\footnote{For a general survey on random graph generation we refer the interested reader to the recent survey by Greenhill~\cite{DBLP:conf/bcc/Greenhill21}.}
This proved to be the first result regarding the mixing time for Curveball Markov Chains.

In this subsection we present the comparison framework that was used in~\cite{DBLP:conf/approx/CarstensK18} and highlight why it is difficult to apply for a comparison of \gesmc with \esmc.
The framework considers a Markov Chain $\mathcal{M}$ and its \emph{heat-bath variant} $\mathcal{M}_{\text{heat}}$ and relates the second largest eigenvalues of both Markov Chains to then compare the mixing times.
For the sake of uniformity we use the same terminology: for an ergodic Markov Chain $\mathcal M = (\Omega, P)$ with stationary distribution $\pi$ where $\pi(x) > 0$ for all $x \in \Omega$, that can be decomposed as
\[ P = \sum_{a \in \mathcal L} \rho(a) \sum_{R \in \mathcal R_a}P_R \] 
where
\begin{enumerate}[i)]
	\item $\mathcal L$ is a finite index set,
	\item $\rho$ a probability distribution over $\mathcal L$,
	\item $\mathcal R_a = \bigcup R_{\ell, a}$ a partition of $\Omega$ for $a \in \mathcal L$
\end{enumerate}
and where the restriction of a matrix $P_R$ to the rows and columns of $R = R_{\ell, a}$ defines the transition matrix of an ergodic, time-reversible Markov Chain on $R$ (and is zero elsewhere), with stationary distribution $\tilde{\pi}_R(x) = \pi(x)/\pi(R)$ for $x \in R$.
As highlighted, a state transition of $\mathcal{M}$ can be thought of as drawing an index $a$ from $\mathcal L$ and then performing a transition on $R$.

Then, the \emph{heat-bath variant} $\mathcal M_{\text{heat}}$ of the Markov chain $\mathcal M$ is given by the transition matrix
\[ P_{\text{heat}} = \sum_{a \in \mathcal L} \rho(a) \sum_{R \in \mathcal R_a} \mathbf{1} \cdot \sigma_R \]
where $\sigma_R$ is a row-vector given by $\sigma_R(x) = \tilde{\pi}_R(x)$ if $x \in R$ and zero otherwise, and $\mathbf{1}$ the all-ones column vector.
Similar to before, $\mathcal M_{\text{heat}}$ can be thought of first drawing an index $a$ from $\mathcal L$ but then simply drawing a state $x$ in $R$ with probability $\tilde{\pi}_R(x)$.

For our purposes, in order to apply this framework we require a suitable index set $\mathcal L$, probability distribution $\rho$ and state space decompositions $\mathcal R_a$ for $a \in \mathcal L$, such that \gesmc is the heat-bath variant $\mathcal M_{\text{heat}}$ of \esmc.
However, in this setting \gesmc must by construction already provide a transition matrix $\mathbf{1} \cdot \sigma_R$ that reflects the stationary distribution $\pi$ up to scaling on all restrictions~$R$ for any $a$.
Naturally and analogously to the original proof in \cite{DBLP:conf/approx/CarstensK18}, the choice for $\mathcal L$ is a set of entries in the adjacency matrix of size $m$ that need to be switched in some order.
In this case, any possible order of execution must be reflected in $\mathcal R_a$ given the chosen entries provided by the initial choice $a$.
Additionally, a transition to any such possible graph must be uniformly given by $\sigma_R$ which in general does not hold.

\section{Parallel Algorithms for \esmc and \gesmc}
\label{sec:steady-algorithm}
In this section, we describe parallel algorithms for \esmc and \gesmc.
Both algorithms rely on \trueparbatch, a parallel algorithm that performs a superstep of edge switches $\sigma_1, \dots, \sigma_{\ell}$ without source dependencies while preserving the observable outcome, i.e. the graph produced is the same as if the switches were performed sequentially.

To this end, we detect all edge switches in $\sigma_1, \dots, \sigma_{\ell}$ that have target dependencies on other switches, and ensure that if a switch $\sigma$ depends on another switch $\sigma'$, then $\sigma$ is decided\footnote{
	We say that a single edge switch is \emph{decided} after we rewire the edges, if it is legal, or reject the switch, if it is illegal.} only after $\sigma'$ is.
For our purposes, it is convenient to think in terms of the following two types of target dependencies:
\begin{itemize}
	\item An edge switch $\sigma_k$ has an \emph{erase dependency} via edge $e$ on switch $\sigma_p$ if $\sigma_k$ has $e$ as target edge, $\sigma_p$ has $e$ as source edge and $k > p$.
	In this case, $\sigma_k$ will attempt to insert an edge that $\sigma_p$ erases, but since $k > p$, it may well be the case that $\sigma_k$ is legal, if $\sigma_p$ is legal, and thus we must ensure that $\sigma_p$ is decided before $\sigma_k$.
	\item An edge switch $\sigma_k$ has an \emph{insert dependency} via edge $e$ on switch $\sigma_q$ if both $\sigma_k$ and $\sigma_q$ have $e$ as target edge and $k > q$.
	In this case, $\sigma_k$ will attempt to insert an edge that $\sigma_q$ inserts, but since $k > q$, it is the case that $\sigma_k$ is illegal, if $\sigma_q$ is legal, and thus we must ensure that $\sigma_q$ is decided before $\sigma_k$.
\end{itemize}

\begin{myobs}\label{obs:target-dependencies}
Given a graph $G=(V, E)$ a superstep of edge switches without source dependencies attempts to remove only edges $e \in E$ and does so only once.
As a direct consequence all erase dependencies for some edge $e \in E$ originate in the same switch $\sigma$.
Also, for any number of insert dependencies for some edge $e$ at most one switch $\sigma$ can be successful.
\hfill $\triangle$
\end{myobs}

\begin{myalgorithm}[ht]
	\KwData{edge list $E$, superstep of switches $S$}
	\SetKwData{undecided}{undecided}
	\SetKwData{delayedSwitches}{delayedSwitches}
	\SetKwData{delay}{delay}
	\SetKwData{illegal}{illegal}
	\SetKwData{stateJ}{stateJ}
	\SetKwData{stateK}{stateK}
	
	$T \gets \emptyset$ \tcp{Initialize dependency table $T$}
	\ParFor{$\sigma_k \in S$}{
		$(i, j, g) \gets \sigma_k$, $e_1 \gets E[i], e_2 \gets E[j]$\;
		Compute target edges $(\vec e_3, \vec e_4) \gets \tau(\vec e_1, \vec e_2, g)$\;
		$\forall e_a \in \{e_1, e_2\}\colon$\ $T$.store($e_a$, $k$, \textsc{erase}, \textsc{undecided})\;
		$\forall e_b \in \{e_3, e_4\}\colon$\ $T$.store($e_b$, $k$, \textsc{insert}, \textsc{undecided})\;
	}
	
	$U \gets S$ \tcp{Initialize array $U$ for undecided switches}
	\While(\CommentSty{// Perform superstep $S$}){$U$ not empty}{
		$D \gets \emptyset$ \tcp{Initialize array $D$ for delayed switches}
		\ParFor{$\sigma_k \in S$}{
				$(i, j, g) \gets \sigma_k$, $e_1 \gets E[i], e_2 \gets E[j]$\;
				Compute target edges $(\vec e_3, \vec e_4) \gets \tau(\vec e_1, \vec e_2, g)$\;
				$s_k \gets \textsc{legal}$ \tcp{Initialize status $s_k$ of $\sigma_k$}
				\For(\CommentSty{// Lookup dependencies}){$e \in \{e_3, e_4\}$}{
					\If{$e$ is self-loop}{
						$s_k \gets \textsc{illegal}$\;
					}
					$p, s_p \gets T$.lookup($e$, \textsc{erase})\;
					$q, s_q \gets T$.lookup\_min($e$, \textsc{insert})\;
					\If{$k < p\ \lor \ s_p = \textsc{illegal}$}{
						$s_k \gets \textsc{illegal}$\;
					}
					\If{$k > q\ \land\ s_q = \textsc{legal}$}{
						$s_k \gets \textsc{illegal}$\;
					}
					\If{$s_k \neq \textsc{illegal}$}{
						\If{${k > p}\ \land\ s_p = \textsc{undecided}$}{
							$s_k \gets \textsc{undecided}$\;
						}
						\If{$k > q\ \land\ s_q = \textsc{undecided}$}{
							$s_k \gets \textsc{undecided}$\;
						}
					}
				}
				
				\If{$s_k = \textsc{legal}$}{
					$E[i] \gets e_3, E[j] \gets e_4$ \tcp{Success, rewire the edges}
				}\ElseIf{$s_k = \textsc{undecided}$}{
					$D$.append($\sigma_k$) \tcp{Delay the switch until the next round}
				}
				$\forall e_a \in \{e_1, e_2\} : $\ $T$.update($e_a$, $k$, \textsc{erase}, $s_k$)\;
				$\forall e_b \in \{e_3, e_4\} : $\ $T$.update($e_b$, $k$, \textsc{insert}, $s_k$)\;
		}
		\textbf{barrier}: wait until all switches completed\;
		$U \gets D$\;
	}
	\caption{\trueparbatch}
	\label{algo:trueparbatch}
\end{myalgorithm}

Algorithm \ref{algo:trueparbatch} shows an implementation of \trueparbatch in pseudocode.
Before performing a superstep, we store the dependencies of the edge switches $\sigma_1, \dots, \sigma_\ell$ in a concurrent hash table $T$.
Then, while attempting to decide a switch, this allows us to lookup the dependencies of the switch and check if the switch is ready to be decided or has unresolved dependencies.
A superstep is then performed incrementally during multiple rounds.
In each round, we only decide switches that have no dependencies on undecided switches.
This in turn resolves the dependencies of all switches that only depend on the decided switches, and as the dependencies cannot be circular, we eventually decide all switches in this way.
Then, finally, once all switches have been decided, the superstep has been performed.

We store the dependencies as tuples in a concurrent hash table, and index them by the source or target edge.
For each switch $\sigma_k$, we store four tuples, one for each source and each target edge, containing the edge $e$, the index $k$ of the switch, the type of operation the switch attempts to perform on the edge $t_{e,k} \in \{\textsc{erase}, \textsc{insert}\}$ and a status flag $s_k \in \{\textsc{undecided}, \textsc{legal}, \textsc{illegal}\}$, that is initially set to $\textsc{undecided}$.
We also use the same data structure to lookup the existence of edges.
To this end, we assume that $T$ implicitly stores a tuple $(e, \infty, \textsc{erase}, \textsc{illegal})$ for each edge that is in the graph, but not a source edge of any switch in the batch, causing a switch that attempts to insert such an edge to be decided as illegal.

Now, when attempting to decide a switch $\sigma_k$, we lookup all tuples where the edge is one of its target edges.
By \cref{obs:target-dependencies}, for each target edge $e$, there is at most one tuple stored by a switch $\sigma_p$ where $t_{e,p} = \textsc{erase}$, i.e. that erases the edge.
Similarly, for each target edge $e$, there is at most one tuple stored by a switch $\sigma_q$, that is legal, and inserts the target edge.
Specifically, at any point, the only tuple that needs to be considered is the tuple with the smallest index $q$ where $t_{e,q} = \textsc{insert}$ and $s_q \neq \textsc{illegal}$.

We then use this information to decide if $\sigma_k$ is legal, illegal, or has to be delayed.
If $k < p$ or $s_p = \textsc{illegal}$, then a target edge of the switch $\sigma_k$ is only erased by a later switch $\sigma_p$, or not erased at all, and thus the switch is illegal.
Similarly, if $k > q$ and $s_q = \textsc{legal}$, then a target edge of $\sigma_k$ is already inserted by the earlier switch $\sigma_q$ and thus the switch is illegal.
Otherwise, if $\sigma_k$ has an erase or insert dependency, but the status of the other switch is $\textsc{undecided}$, the switch cannot be decided yet, and must be delayed.
In any other case the switch is legal.

If the switch is legal, the edges are rewired and the status of the tuples is set to $\textsc{legal}$.
Otherwise, if the switch is illegal, the status of the tuples is set to $\textsc{illegal}$.
Finally, if the switch has unresolved dependencies, we delay it until the next round.

\subsection{ParES}

\begin{myalgorithm}[!t]
	\KwData{edge list $E$, requested number of switches $r$}
	
	\tcp{Initialize array of requested switches $R$}
	$R \gets \emptyset$\;
	\ParFor{$k$ from $1$ to $r$}{
		Sample edge indices $i, j \sim [m]$ with $i \neq j$\;
		Sample direction bit $g \sim \{0, 1\}$\;
		$R[k] \gets (i, j, g)$\;
	}
	
	$s \gets 1$\;
	\While{$s \leq r$}{
		\tcp{Initialize hash set $H$}
		$H \gets \emptyset$\;
		\tcp{Find superstep without source dependencies}
		$t \gets r + 1$\;
		\ParFor{$k$ from $s$ to $(t - 1)$}{
			\tcp{Check for collision and update lower bound $t$}
			$(i, j, g) \gets R[k]$\;
			$k' \gets H$.insert\_if\_min$(i, k)$\;
			 $k'' \gets H$.insert\_if\_min$(j, k)$\;
			$t' \gets \max \{k, k'\}$, $t'' \gets \max \{k, k''\}$\;
			$t \gets \min \{t, t', t''\}$\;
		}
		\tcp{Perform superstep $\sigma_s, \dots, \sigma_{t-1}$}
		$S[k - s + 1] \gets R[k] \quad \forall s \leq k \leq t - 1$\;
		\textsc{ParallelSuperstep}$(E, S)$\;
		$s \gets t$\;
	}
	\caption{\truepares}
	\label{algo:truepares}
\end{myalgorithm}

We first describe how \truepares, a parallelization of \esmc, can be implemented by using \trueparbatch (see Algorithm \ref{algo:truepares}).

The algorithm first populates an array $R$ with the requested number of switches $r$ by sampling two edge indices and a direction bit for each switch.
The switches in $R$ are then performed during multiple iterations, where each iteration performs a superstep.
In each iteration, we let $s$ denote the number of switches that were already performed.
We then identify the next superstep by finding the smallest index $t > s$ of a switch that has a source collision with another switch in the sequence $\sigma_{s}, \dots, \sigma_r$.
To this end, we insert for each switch $\sigma_k$ two tuples $(i, k)$ and $(j, k)$ into a concurrent hash set.
If for any edge index $i$, a different switch $\sigma_{k'}$ exists which has inserted the same index $i$, we know that the first collision occurs at most at index $t' = \max \{k, k'\}$ so we set the lower bound to $t \gets \min \{t,  t'\}$.
In addition, if $k < k'$, we replace the tuple $(i, k')$ by $(i, k)$ to detect further switches $k < k'' < k'$ which may update the lower bound.
Once all switches up to the lower bound $t$ have been inserted, we know that $\sigma_{s}, \dots, \sigma_{t-1}$ is a sequence of edge switches without source dependencies.
These edge switches are then performed by using \trueparbatch.

\subsection{ParGlobalES}

\begin{myalgorithm}[!t]
	\KwData{edge list $E$, requested number of global switches $r$}
	
	\For{$i$ from $1$ to $r$}{
		\tcp{Select random global switch $\Gamma = (\pi, \ell)$}
		Sample random permutation $\pi$ of $[m]$\;
		Sample $\ell \sim Binom(\lfloor m / 2 \rfloor, 1 - P_L)$\;
		\tcp{Perform global switch}
		$S[k] \gets (\pi(2k - 1),\, \pi(2k), \, \1_{\pi(2k - 1) < \pi(2k)}) \quad \forall 1 \leq k \leq \ell$\;
		\textsc{ParallelSuperstep}$(E, S)$\;
	}
	\caption{\trueparglobes}
	\label{algo:trueparglobes}
\end{myalgorithm}

Implementing the parallelization \trueparglobes of \gesmc is much simpler (see Algorithm \ref{algo:trueparglobes}).
As a global switch $\Gamma$ contains no source dependencies by definition, the full algorithm only consists of one loop, which in each iteration selects a random global switch $\Gamma = (\pi, \ell)$ and then calls on \trueparbatch to perform this switch.

\subsection{Analysis of ParGlobalES}

A necessary condition for good scaling of \trueparglobes is that the number of rounds to perform a global switch is small.
Note that in practice, the number of rounds depends on the scheduling and assignment of switches to processors.
This is because we store the dependencies in a concurrent hash table: if a switch $\sigma$ depends on a switch $\sigma'$, but the processor assigned to $\sigma'$ finishes writing the result before the processor assigned to $\sigma$ attempts to decide its switch, then both switches can be decided in the same round, and the dependency will not affect the number of rounds.
Similarly, if $\sigma$ and $\sigma'$ are assigned to the same processor, the dependency will not affect the number of rounds.
Still, even for a worst-case scheduler, that always moves dependent switches to the next round, we can show that the expected number of rounds is small for bounded degree graphs, regular graphs, and power-law graphs with sufficiently high degree exponent.

\begin{mythm} \label{thm:std-g-es-scales}
	Let $R$ denote the number of rounds needed to perform a global switch $\Gamma = (\pi, \ell)$ on a graph $G = (V, E)$.
	If $G$ has $m = |E|$ edges and each node has at most degree $d \leq \Delta$, we have $R \leq 4 \Delta^2 / m$ in expectation over $\pi$.
\end{mythm}

\begin{mycol} \label{col:d-bound}
	If each node in $G$ has at most degree $d \leq \sqrt{m}$, we have $R \leq 4$ in expectation over $\pi$.
\end{mycol}

\begin{mycol} \label{col:d-bound}
	If $G$ is a $d$-regular graph, we have $R \leq 4$ in expectation over $\pi$.
\end{mycol}

\begin{proof}
	Observe that $R < k$ unless there exists a chain of switches $\sigma^{(1)}, \dots, \sigma^{(k)} \in \Gamma$, so that each switch $\sigma^{(r+1)}$ with $1 \leq r < k$ depends on switch $\sigma^{(r)}$ and cannot be decided before round $r + 1$.
	There are four ways that $\sigma^{(r+1)}$ can depend on $\sigma^{(r)}$, either (1) $\sigma^{(r)}$ erases edge $e_a$, and $\sigma^{(r+1)}$ inserts $e_a$, or (2) $\sigma^{(r)}$ inserts edge $e_b$, and $\sigma^{(r+1)}$ inserts $e_b$, or (3) and (4), the symmetric cases for the other two edges inserted or erased by $\sigma^{(r)}$.
	Recall that a switch $\sigma_i \in \Gamma$ inserts edge $e = \{u, v\}$ if the permutation $\pi$ contains an edge incident with $u$ and an edge incident with $v$ in positions $2i - 1$ and $2i$.
	There are $d_u$ and $d_v$ such edges, respectively, and as only one assignment of the edges to the positions will give the necessary direction bit $g_i$, each edge must be assigned to one specific position.
	Thus, the probability that switch $\sigma_i$ inserts edge $e$ is at most
	\begin{equation}
		\frac{\overbrace{d_u d_v (m - 2)!}^{\text{\# permutations such that $\sigma_i$ inserts $e$}}}{\underbrace{m!}_{\text{\# total permutations}}} = \frac{d_u d_v}{m (m - 1)} \leq \frac{\Delta^2}{m (m - 1)} \leq \frac{2 \Delta^2}{m^2}.
	\end{equation}
	Hence, for each switch $\sigma_i \in \Gamma$, the probability that the switch with the next index $\sigma_{i+1}$ inserts one of the four edges, and depends on $\sigma_i$, is at most $8 \Delta^2 / m^2$.
	Then, in expectation, if switch $\sigma^{(r)}$ has index $i$, $\sigma^{(r+1)}$ has index at least $i + m^2 / 8 \Delta^2$, and since $\Gamma$ contains only $m / 2$ switches, each chain has expected length $k \leq 4 \Delta^2 / m$.
\end{proof}

The following theorem gives a sharper bound for graphs with a skewed degree sequence such as power-law graphs.

\begin{mythm} \label{thm:std-g-es-scales2}
	Let $R$ denote the number of rounds needed to perform a global switch $\Gamma = (\pi, \ell)$ on a graph $G = (V, E)$, let $S = \{ \{u, v\} : u, v \in V, u \neq v\}$ denote the set of possible edges, and let $P_2 = \sum_{e = \{u, v\} \in S} (d_u d_v / m (m-1))^2$.
	Then, we have $R = \Oh{P_2 m}$ in expectation over $\pi$.
\end{mythm}

\begin{proof}
	For each switch $\sigma_i \in \Gamma$, the probability that the switch with the next index $\sigma_{i+1}$ inserts a specific edge $e$ that $\sigma_i$ erases is
	\begin{equation}
		\sum_{e = \{u, v\} \in S} \underbrace{\frac{\1_{e \in E} }{m}}_{\text{$\sigma_i$ erases $e$}} \underbrace{\frac{d_u d_v}{m (m-1)}}_{\text{$\sigma_{i+1}$ inserts $e$}} \leq \frac{1}{m} \sum_{e = \{u, v\} \in S} \frac{d_u d_v}{m (m-1)} \leq \frac{1}{m}.
	\end{equation}
	For an insert dependency to arise, $\sigma_{i+1}$ must insert a specific edge $e$ that $\sigma_i$ inserts, and the probability for this event is given by
	\begin{equation}
		\sum_{e = \{u, v\} \in S} \underbrace{\frac{d_u d_v}{m (m-1)}}_{\text{$\sigma_{i}$ inserts $e$}} \underbrace{\frac{(d_u - 1) (d_v - 1)}{(m - 2) (m - 3)}}_{\text{$\sigma_{i+1}$ inserts $e$}} = \Oh{P_2}.
	\end{equation}
	Hence, for each switch $\sigma_i \in \Gamma$, the probability that the switch with the next index $\sigma_{i+1}$ depends on $\sigma_i$ is at most $2 / m + \Oh{P_2}$, and as $\Gamma$ contains $m / 2$ switches, each dependency chain has expected length $\Oh{P_2 m}$.
\end{proof}

\begin{mylem}
	If $G$ is a power-law graph with degree exponent $\gamma > 2$, we have $P_2 = \Oh{1 / n + n^{(2 + \gamma - \gamma^2)/(\gamma - 1)}}$.
\end{mylem}

\begin{mycol}
	If $G$ is a power-law graph with degree exponent $\gamma > 1 + \sqrt{2}$, we have $R = \Oh{1}$ in expectation over $\pi$.
\end{mycol}

\begin{proof}
	We will use known properties of power-law graphs to find a suitable upper bound on $P_2$.
	Recall that if $G$ is a power-law graph with degree exponent $\gamma > 2$, then $G$ has $N_d = \Oh{n d^{-\gamma}}$ nodes with degree $d$ and maximum degree $\Delta = \Oh{n^{1/(\gamma -1)}}$.
	Rewriting the expression for $P_2$ as a sum over degree groups yields
	\begin{align}
		P_2 &= \sum_{i = 1}^{\Delta} N_i \sum_{j = 1}^{\Delta} N_j \left( \frac{i j}{m (m - 1)} \right)^2
		\\
		&\leq \sum_{i = 1}^{\Delta} N_i \sum_{j = 1}^{\Delta} N_j \frac{i^2 j^2}{n^4} 
		\\
		&= \Oh{ \sum_{i = 1}^{\Delta} \frac{1}{i^\gamma} \sum_{j = 1}^{\Delta} \frac{1}{j^\gamma} \frac{i^2 j^2}{n^2} }.
	\end{align}
	Observe that the terms in $P_2$ where $i j < n$ sum to at most $\Oh{1 / n}$.
	For each of the remaining terms, we have $i j \geq n$, so we may write their sum as
	\begin{equation}
		\sum_{i = 1}^{\Delta} \frac{1}{i^\gamma} \sum_{j = \lfloor \frac{n}{i} \rfloor}^{\Delta} \frac{1}{j^\gamma} \frac{i^2 j^2}{n^2}.
	\end{equation}
	Now, using $j^{\gamma} > j^2$, and substituting $\lfloor \frac{n}{i} \rfloor$ for $j$, we see that $i^\gamma / n^\gamma$ is an upper bound on each term of the inner sum over $j$, and obtain
	\begin{equation}
		\sum_{i = 1}^{\Delta} \frac{1}{i^\gamma} \sum_{j = \lfloor \frac{n}{i} \rfloor}^{\Delta} \frac{1}{j^\gamma} \frac{i^2 j^2}{n^2} \leq \sum_{i = 1}^{\Delta} \frac{1}{i^\gamma} \sum_{j = \lfloor \frac{n}{i} \rfloor}^{\Delta} \frac{i^\gamma}{n^{\gamma}} \leq \frac{\Delta^2}{n^{\gamma}} = \Oh{n^{(2 + \gamma - \gamma^2)/(\gamma - 1)}}.
	\end{equation}
\end{proof}

\section{Implementation}
\label{sec:implementation}
In this section, we describe the implementation of our sequential and parallel algorithms and the data structures used therein.
In addition to the \trueparglobes algorithm described in \cref{sec:steady-algorithm}, we implement the following sequential and parallel algorithms.
\begin{itemize}
	\item \fastes: a fast sequential implementation of \esmc.
	\item \globes: a sequential implementation of \gesmc.
	\item \pares: a simplistic parallelization of \esmc.
\end{itemize}
All algorithms (including previously existing ones) are implemented in C++.

\subsection{NaiveParES}

To establish a performance baseline for parallel algorithms, we implement \pares, a simplistic parallelization of \esmc.

Each PU (processing unit) performs switches independently while synchronizing implicitly only by preventing concurrent updates of individual edges.
To ensure that no edge is erased or inserted twice, we store the edges in a concurrent hash-set using the following semantics:
to remove an edge from the set, a ticket has to be acquired first; this can be done by locking an existing edge or by inserting-and-locking a new edge.
These operations are implemented using a compare-and-exchange primitive.
Concurrent updates to the same edge are sequenced by the hardware.

Note that this implementation performs all edge switches that are legal after synchronization but ignores dependencies between edge switches.
In contrast to the exact parallelizations \truepares and \trueparglobes, it can thereby deviate from the intended Markov Chain.

\subsection{Graph and dependency representation}\label{subsec:impl-hash-set}
Most \esmc implementations use an adjacency list to store the graph and manipulate it with each switching~\cite{DBLP:journals/jpdc/BhuiyanKCM17,DBLP:journals/netsci/StaudtSM16,SciPyProceedings_11,Lancichinetti2009,DBLP:journals/corr/abs-cs-0502085}\footnote{
	\cite{DBLP:journals/jpdc/BhuiyanKCM17} use a \emph{reduced} adjacency matrix that only stores one directed edge.
	\cite{Lancichinetti2009} use a different MC with the same switchings.
	\cite{DBLP:journals/corr/abs-cs-0502085} interleave \esmc with connectivity checks after each edge switching. They use an adjacency list where high-degree nodes store their neighborhoods in individual hash tables.
}.
This design choice often leads to an easy integration with other algorithms.
However, \esmc requires a graph representation that efficiently supports edge insertion, deletion, and existence queries.
Unfortunately, an adjacency list cannot support updates and search both in constant time (cf. the hybrid data structure of \cite{DBLP:journals/corr/abs-cs-0502085}).

In contrast, hash-sets support all required operations in expected constant time.
To this end we first identify each possible edge with a unique integer.
For instance, if nodes $u, v \in V$ are stored as 32-bit integers, then each possible edge $\{u, v\} \in E$  can be identified by the 64-bit integer where the first 32-bits are set equal to the smaller node $u < v$ and the remaining 32-bits are set equal to the larger node $v > u$ (recall that we disallow loops).
To store a graph in a hash set we insert for each edge its unique identifier.
Checking if an edge is contained in the graph is possible by checking if its identifier is contained in the hash-set.
When performing a legal edge switch, we delete both identifiers of the source edges and then insert both identifiers of the target edges.

From a practical point of view, we require a hash-set implementation that can handle a roughly balanced mix of insertions, deletions, and search queries.
After preliminary experiments on various graphs, machines, and hash-set implementations\footnote{
	We considered the following hash-sets: \url{https://gcc.gnu.org/},
	\url{https://github.com/\{Tessil/robin-map, Tessil/hopscotch-map, sparsehash/sparsehash\}}.
	
},
we find that in most cases \textsc{RobinMap} with a maximum load-factor of $1/2$ is the fastest sequential solution.
Observe that for performance reasons, all our implementations use hash-tables where the number of buckets is a power-of-two; hence the actual load-factor can be lower.
Our hash function uses the 64bit variant of the \texttt{crc32} instruction available on \texttt{x64} processes with SSE 4.2~\cite{intel-manual-19}.

Our parallel algorithms require concurrent hash-tables with stable iterators (\ie once an element is placed into a bucket, it is not moved until it gets erased).
It is folklore that such a data structure can be efficiently implemented with open addressing  and lock-free compare-and-swap instructions (cf. \cite{DBLP:journals/topc/MaierSD19}).

We implement the locking of edges as follows: each edge is kept in an 64bit-wide bucket, where 56 bits are used to store the edge and 8 bit are reserved for locking.
To acquire a lock, a PU tries to compare-and-swap its thread id into the lock bits and succeeds only if the bucket previously kept the edge in an unlocked state.
This implementations allows us to process graphs with up-to $n \leq 2^{28}$ nodes on $\nproc < 256$ threads.
Observe that these restrictions can be lifted quite easily, as virtually all relevant processors support 128bit compare-and-swap instructions with only moderate performance penalties.

\subsection{Sampling edges}
Pseudo-random bits are generated using the MT19937-64 variant of the Mersenne Twister~\cite{DBLP:journals/tomacs/MatsumotoN98} implemented by \texttt{libstdc++} and translated into unbiased random integers using \cite{DBLP:journals/tomacs/Lemire19}.
Random permutations are sampled in parallel with an optimized implementation inspired by~\cite{DBLP:journals/ipl/Sanders98,DBLP:journals/tomacs/Lemire19}.

To sample edges uniformly at random we consider two options:
Firstly, we maintain an auxiliary array of edges. In order to sample an edge uniformly at random, we read from a random index --- this closely resembles the way we introduced edge switching in the previous chapters. Secondly, the use of open-addressing hash-tables allows us to directly sample from the hash-set by repeatedly drawing random buckets until we hit the first non-empty one.

While the second option avoids additional memory, it leads to a time trade-off: while all queries discussed in \cref{subsec:impl-hash-set} benefit from a low load-factor $L$, the sampling time is geometrically distributed with a success probability of $L$.
It, thus, favors a high load-factor.
In preliminary experiments, we found that decreasing the load factor to allow faster queries and sampling using an additional array yields up-to 30~\% faster overall performance compared to balancing the load factor for both queries and sampling.

\subsection{Prefetching}
The rewiring of random edges inherently leads to unstructured accesses to main memory, especially if the graph is represented in a hash set.
While \cite{DBLP:journals/jea/HamannMPTW18} propose an I/O-efficient edge switching implementation for graphs exceeding main memory, their solution requires to repeatedly sort the edge list (and other data structures).
In the context of a parallel algorithm, this sorting step alone is more expensive than a global switch using unstructured accesses.

We therefore accept the random I/Os and accelerate them using prefetching instructions.
To this end, we split all insertion, deletion, and search queries to our hash-sets into two:
in a first step, we hash the key and identify the bucket in which the item is placed if there is no collision.
We then prefetch this bucket as well as its direct successor, and return precomputed values that are required when we carry out the actual operation in a second step.
Since we use linear probing hash-sets with a low load factor and a prefetch in advance, we effectively eliminate almost all cache misses if there is sufficient time between both steps.
To increase this time window, we use a pipeline of four edge switches in different progress stages.

\section{Experiments}
\label{sec:experiments}
In the following, we empirically investigate the mixing times of the Markov Chains and the runtime performances of their derived algorithms.
The performance benchmarks are built with GNU g++-9.3 and executed on a machine equipped with an AMD EPYC 7702P 64-core processor running Ubuntu 20.04.

Our experiments are performed on the following datasets:
\begin{itemize}
	\item \textsc{(SynGnp)} --- We generate $G(n,p)$ graphs~\cite{Gilbert59} (each edge exists independently with probability~$p$) for varying node counts $n$ and $p$.
	\item \textsc{(SynPld)} --- For varying node counts $n$ and degree exponents $\gamma$, we generate power-law degree sequences according to the degree distribution $\textsc{Pld}([1..\Delta], \gamma)$
	 where the maximum degree is set to $\Delta = n^{1/(\gamma - 1)}$ matching the analytic bound~\cite{DBLP:conf/soda/GaoW18}.
	Thereafter, the generated sequences are materialized by the Havel-Hakimi algorithm \cite{Havel1955, doi:10.1137/0110037}.
	Both steps are performed using NetworKit~\cite{DBLP:journals/netsci/StaudtSM16}.
	\item \textsc{(NetRep)} ---	We consider graphs from the network repository~\cite{DBLP:conf/aaai/RossiA15} where we exclude the unsuitable categories \emph{dimacs}, \emph{dimacs10}, \emph{graph500} (benchmark graphs), \emph{dynamic}, \emph{misc} (unclassified graphs), \emph{rand} (synthetic graphs) and \emph{tscc} (temporal graphs). To ensure that the graphs are simple and undirected, we perform the following modifications: all directed edges $(u, v)$ are replaced by the undirected edges $\{u, v\}$, and self-loops and multi-edges are removed.
\end{itemize}

\subsection{Empirical Mixing Time of \esmc and \gesmc}
\label{subsec:mixing}

\begin{figure}[!t]
	\centering
	\includegraphics[width=\figscale \linewidth]{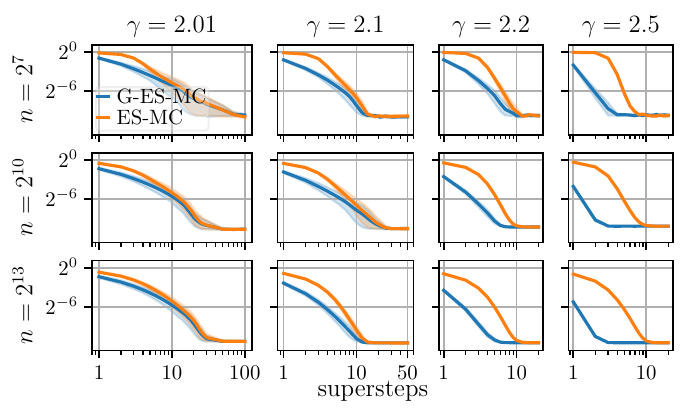}
	\caption{
		Fraction of non-independent edges as a function of the thinning value $k$ as a multiple of the supersteps for the \textsc{SynPld} dataset where $(n, \gamma) \in \{2^7,2^{10},2^{13}\}\times\{2.01, 2.1, 2.2, 2.5\}$.
		Each data point is represented by its mean value $\mu$ (line) and its $2\sigma$ error (shade).
	}
	\label{fig:autocorrelation-synthetic-powerlaw-lineplot}
\end{figure}

\begin{figure}[!t]
	\centering
	\includegraphics[width=\figscale \linewidth]{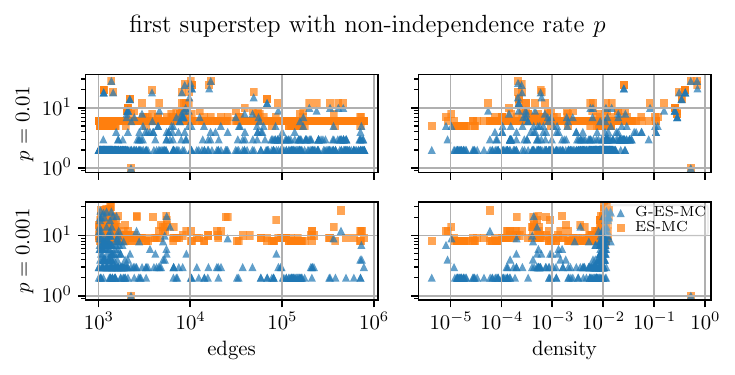}
	\caption{
		Scatterplots the \textsc{NetRep} dataset where the $x$-coordinate either denotes the number of edges $m$ (left) or the density $m/\binom{n}{2}$ (right) of a graph with $n$ nodes and $m$ edges.
		The $y$-coordinate represents the first superstep $k$ at which the mean fraction of non-independent edges drops below either \SI{1e-2}{} (top) or \SI{1e-3}{} (bottom).
		We do not use large primes and numbers with many divisors as thinning values.
		This yields an uneven (but inconsequential) quantization of the y-axis.
		The outlier that merely requires $k = 1$ supersteps for both Markov Chains possesses only two unique node degrees.
	}
	\label{fig:autocorrelation-realworld-scatterplot}
\end{figure}

Here, we compare the mixing times of \esmc and the novel \gesmc.
While we argue in \cref{sec:g-es-mc} that the lack of source dependencies of \gesmc improves parallelizability over \esmc, it is a priori unclear whether this restriction affects the randomization quality of the Markov Chain.

In practice, the mixing time is approximated by empirical proxies or estimated by data driven methods~\cite{DBLP:journals/jea/StantonP11, DBLP:journals/compnet/RayPS15}.
The former measures the convergence to the stationary distribution by convergence of its proxy or by some aggregated value.
In doing so, the Markov Chain is reflected by a projection which may converge faster~\cite{berger2018smaller}.
The result depends on the proxy and might be insufficient for other more sensitive proxies.
Additionally, it has been observed that common measures, \eg~assortativity coefficients, clustering coefficients, diameter, maximum eigenvalue and triangle count, are less sensitive than data-driven methods~\cite{DBLP:journals/jea/HamannMPTW18, DBLP:journals/compnet/RayPS15}.
Thus, we consider the \emph{autocorrelation analysis}, an approximate non-parametric method \cite{DBLP:journals/compnet/RayPS15}.

The autocorrelation analysis proceeds in two steps.
First, execute the Markov Chain for a large number of steps $K$, and for each possible edge~$e$ track in a binary time-series $\{Z_t\}$ whether edge~$e$ exists at time~$t$.
By its own, $\{Z_t\}$ and its transitions will be correlated \cite{DBLP:journals/jea/StantonP11} which naturally indicates that a single Markov Chain step is insufficient.

Consider now the $k$-thinned chain $\{Z^k_t\}$ which retains every $k$-th entry of $\{Z_t\}$.
The $k$-thinned chain will have smaller autocorrelation and begin to resemble independent draws from a distribution for sufficiently large $k$.
At some point, the thinned time-series should resemble an independent process more than a first-order Markov process.
To determine which of the models is a better fit, the Bayesian Information Criterion (BIC) is computed using the $G^2$-statistic~\cite{bishop2007discrete} (see~\cite{DBLP:journals/compnet/RayPS15} for details).
Thus, in a second step, the time-series $\{Z_t\}$ is progressively thinned to determine independent edges for the thinned time-series $\{Z^k_t\}$ for increasing values of $k$.

For our purposes, instead of first computing the whole time-series $\{Z_t\}$ and then considering increasing thinning values in a post-processing step, we define a fixed set of thinning values $T$ and aggregate relevant entries of $\{Z_t\}$ on-the-fly for each $k \in T$.
While this approach is far less memory-consuming, we cannot recover for each edge the earliest point of time it would have been deemed independent.
Instead, for a thinning value $k$, we report the fraction of edges that would be deemed independent irrespective of a smaller thinning value $k' < k$.

In this context, we compare \esmc and \gesmc.
In order to visually align the results we define a \emph{superstep} for both Markov Chains.
To this end, let $m/2$ uniform random edge switches and one uniform random global switch be a designated superstep for \esmc and \gesmc, respectively.
This accounts for the fact that one global switch potentially executes $m/2$ (non-uniform) edge switches.

We first consider \textsc{SynPld} and generate for each $(n, \gamma) \in \{2^7, 2^{10}, 2^{13}\} \times \{2.01, 2.1, 2.2, 2.5\}$ forty power-law graphs (we limit the largest node count to $n = 2^{13}$ since the longest individual run already took $18$ hours using an Intel Skylake Gold 6148 processor).
In \cref{fig:autocorrelation-synthetic-powerlaw-lineplot}~we report the mean fraction of non-independent edges depending on the number of supersteps for a subset of the node counts and degree exponents.
For highly skewed degree sequences, \eg~$\gamma = 2.01$, the  \gesmc performs slightly better than the  \esmc for small supersteps.
Increasing the number of supersteps results in matching performances for both.
For larger degree exponents $\gamma \ge 2.2$ \gesmc consistently outperforms \esmc where the advantage increases with $\gamma$.
We observe both features for up to two orders of magnitude, and we expect this to hold for even larger values of $n$.

Next we investigate real-world graphs of the \textsc{NetRep} dataset.
Due to the high computational cost, we restrict ourselves to graphs with $\num{1000} \le m \le \num{800000}$ edges.
To further reduce the cost, we perform the autocorrelation analysis only for the edges of the initial graph, reducing the memory footprint of each thinning to $\Theta(m)$ where $m$ is the number of edges.
In \cref{fig:autocorrelation-realworld-scatterplot}~we present for each graph the first reported superstep at which the mean fraction of non-independent edges of at least $15$ runs is below a threshold $\tau$.
For $\tau = 1 \times 10^{-2}$, \gesmc seems to consistently outperform \esmc except for very dense graphs where the performance is similar.
The $\tau = 1 \times 10^{-3}$ is reached by only $46$\% of the $594$ instances within $30$ supersteps.
Here, \gesmc still outperforms \esmc on most instances except for moderately dense graphs on which both chains converge significantly slower.

\subsection{Performance Benchmarks}

\begin{figure}[!t]
	\scriptsize
	\renewcommand{\arraystretch}{1.5}
	\caption{
		\normalfont Runtimes in seconds on a sample of graphs from \textsc{NetRep} sorted by network size.
		The left columns lists the graph, number of nodes $n$, number of edges $m$ and maximum degree $d_{\max}$.
		The center columns list the sequential and parallel implementations for $\nproc = 1$ PU.
		The right columns list the parallel implementations for $\nproc = 32$ PUs. 
		The best time in each group is indicated by the bold font.
		A dash (---) indicates a runtime of more than $1000$ sec. for $20$ supersteps.}
	\centering
	
	\begin{condRotTab}
  \rowcolors{2}{gray!20}{white}
	\begin{tabular}{ p{0.15\textwidth} p{0.06\textwidth}<{\raggedleft} p{0.07\textwidth}<{\raggedleft} p{0.06\textwidth}<{\raggedleft} | p{0.03\textwidth}<{\centering} p{0.03\textwidth}<{\centering}  p{0.03\textwidth}<{\centering}  p{0.03\textwidth}<{\centering} |  p{0.03\textwidth}<{\centering}  p{0.03\textwidth}<{\centering} |  p{0.03\textwidth}<{\centering} p{0.03\textwidth}<{\centering}  p{0.03\textwidth}<{\centering}  p{0.03\textwidth}<{\centering} }
		& & & & \diagonal{NetworKit} & \diagonal{Gengraph} & \diagonal{\fastes} & \diagonal{\globes} & \diagonal{\pares} & \diagonal{\textsc{ParGlobalES}} & \diagonal{\pares} & \diagonal{\textsc{ParGlobalES}} \\  \hline 
		Graph & $n$ & $m$ & $d_{\max}$ & \multicolumn{6}{c|}{$\nproc=1$} & \multicolumn{2}{c}{$\nproc=32$}\\ \hline
		soc-twitter-mpi-sws & $41$ M & $1.2$ B & $2.9$ M & --- & --- & --- & --- & --- & --- & $\mathbf{251}$ & $397$  \\ 
		bn-human-Jung2015 & $1.8$ M & $146$ M & $8.7$ K & --- & --- & $517$ & $\mathbf{460}$ & $\mathbf{448}$ & $784$ & $\mathbf{20.0}$ & $36.7$ \\ 
		tech-p2p & $5.7$ M & $140$ M& $675$ K & --- & --- & $530$ & $\mathbf{464}$ & $\mathbf{477}$ & $788$ & $\mathbf{21.3}$ & $37.2$ \\ 
		socfb-konect & $59$ M & $92$ M & $4.9$ K & --- & --- & $287$ & $\mathbf{253}$ & $\mathbf{228}$ & $459$ & $\mathbf{11.9}$ & $21.7$ \\ 
		ca-holywood2009 & $1$ M & $56$ M & $11$ K & --- & $686$ & $140$ & $\mathbf{112}$ & $\mathbf{116}$ & $244$ & $\mathbf{8.1}$ & $11.4$ \\ 
		inf-road-usa & $23$ M & $28$ M & $9$ & $619$ & $186$ & $49.2$ & $\mathbf{41.4}$ & $\mathbf{53.6}$ & $97.0$ & $5.2$ & $\mathbf{5.1}$ \\ 
		bio-human-gene1 & $220$ K & $12$ M & $7.9$ K &$512$ & $109$ & $\mathbf{12.3}$ & $12.5$ & $\mathbf{18.1}$ & $32.0$ & $\mathbf{1.3}$ & $2.0$ \\ 
		web-wikipedia2009 & $1.8$ M & $4.5$ M & $2.6$ K & $65.4$ & $36.5$ & $\mathbf{4.7}$ & $4.9$ & $\mathbf{6.6}$ & $9.7$ & $\mathbf{0.58}$ & $0.95$  \\ 
		cit-HepTh & $22$ K & $2.4$ M & $8.7$ K & $45.0$ & $20.4$ & $\mathbf{2.2}$ & $2.3$ & $\mathbf{3.4}$ & $5.3$ & $\mathbf{0.25}$  & $0.47$ \\ 
		email-enron-large & $33$ K & $180$ K & $1.3$ K & $0.92$ & $0.44$ & $\mathbf{0.12}$ & $0.14$  & $\mathbf{0.21}$ & $0.37$ & $\mathbf{0.06}$ & $0.07$ \\ 
		rec-amazon & $91$ K & $120$ K & $5$ & $0.57$ & $0.16$ & $\mathbf{0.10}$ & $0.11$ & $0.18$ & $\mathbf{0.17}$ & $0.06$ & $\mathbf{0.05}$ \\ 
	\end{tabular}
	\label{table:rt-netrep-sample}
	\end{condRotTab}
\end{figure}

In this section, we benchmark our \esmc and \gesmc implementations.
In each experiment, we run a subset of the implementations on the same initial graph and measure the average time required to initialize the data structures and perform $20$ supersteps (e.g. $10$ switches per edge).
In practice, common  choices~\cite{Milo2003,DBLP:conf/alenex/GkantsidisMMZ03,DBLP:conf/waw/RayPS12} are $10$ to $30$ switches per edge.
As \gesmc typically requires fewer supersteps (compare \cref{subsec:mixing}), this gives a slight advantage to \esmc over the \gesmc implementations.

\subsubsection{Runtime}

We compare existing sequential implementations to our solutions and report absolute runtimes.
To this end, we benchmark all implementations on a sample of graphs from \textsc{NetRep} and report their runtimes in Table \ref{table:rt-netrep-sample}.
We select the graphs in this sample to cover a variety of sizes, average degrees and maximum degrees.
As some of the networks are quite large, we set a timeout of $1000$ seconds.

We first compare \fastes and \globes, our sequential \esmc and \gesmc solutions, with existing implementations from NetworKit \cite{DBLP:journals/netsci/StaudtSM16} and Gengraph \cite{DBLP:journals/compnet/VigerL16}.
Our solutions run 15-50 times faster than NetworKit and 5-10 times faster than Gengraph.
We also observe that \globes is faster than \fastes on large graphs, where shuffling is more efficient than sampling the edges, whereas \fastes runs faster on small graphs.
In conclusion, our sequential implementations provide a meaningful baseline to measure further speed-ups.

Next, we report the runtimes of the parallel algorithms.
For $\nproc=32$ PUs, all parallel implementations run much faster than the sequential implementations.
On the largest graph, only the parallel implementations were able to perform $20$ supersteps before the timeout.
Here, \trueparglobes is up to $12$ times faster than \globes.
On all graphs, \trueparglobes only shows a slowdown of at most $2$ compared to the baseline algorithm \pares.
In this context, it is important to recall that \pares is not an exact parallelization since it ignores dependencies between edge switches.

\begin{figure}[t]
 \centering
 \includegraphics[width=0.8\textwidth]{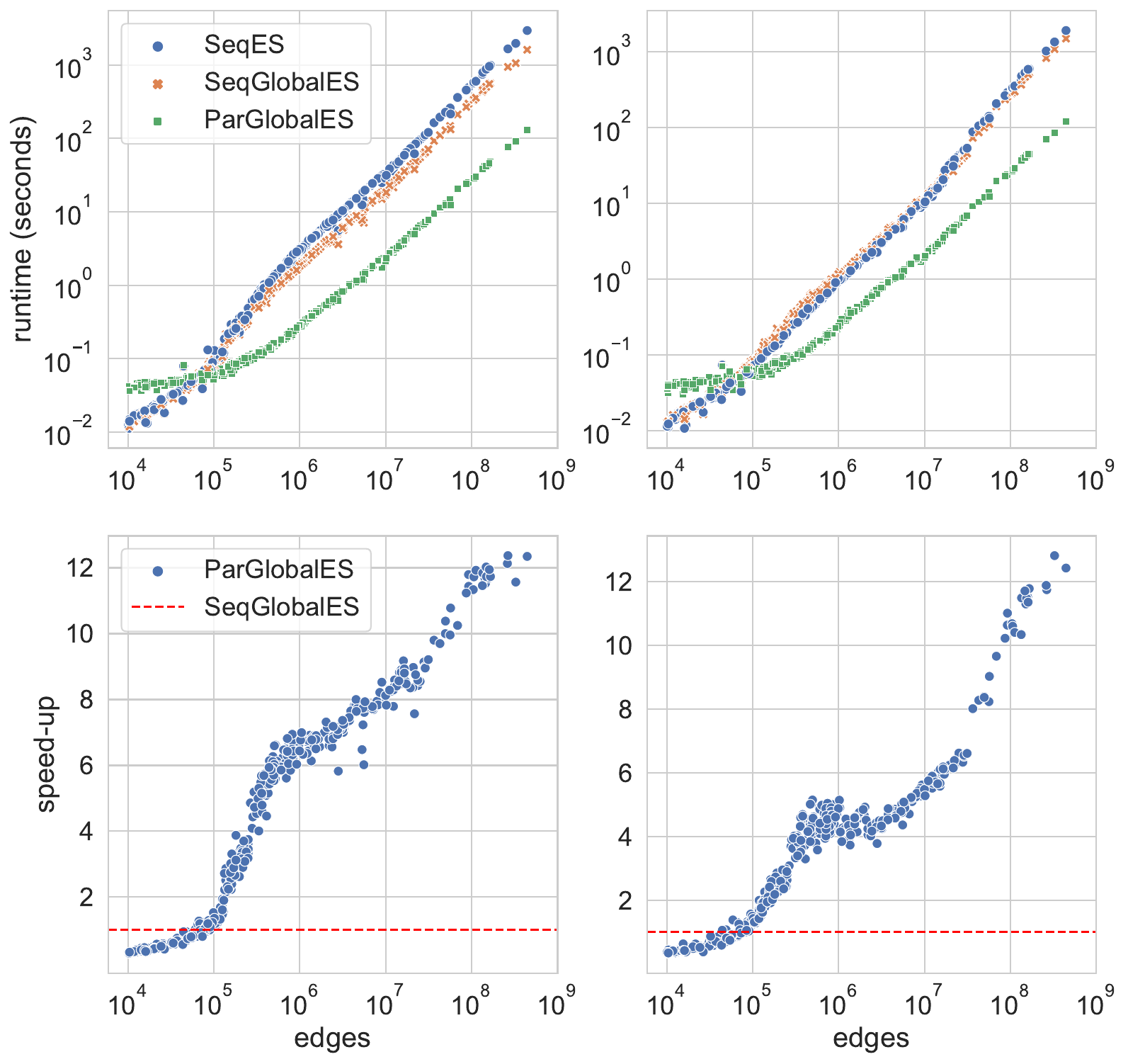}
 \caption{Scatterplot showing runtimes and speed-ups on \textsc{NetRep}. Top row: runtimes of \fastes and \globes on $\nproc=1$ PU and \trueparglobes on $\nproc=32$ PUs. The color and symbol indicates the algorithm. Bottom row: speed-ups of \trueparglobes over \globes. The height of the red dashed line corresponds to a speed-up of $1$. Left column: without prefetching. Right column: with prefetching.}
 \label{fig:real-world_scatter}
\end{figure}

In \cref{fig:real-world_scatter}, we evaluate \fastes, \globes and \trueparglobes on all graphs from \textsc{NetRep} with at least $m = 10^4$ edges.
For each graph, we run \fastes and \globes on one PU and \trueparglobes on $\nproc=32$ PUs and report the absolute runtimes and the speed-up of \trueparglobes over \globes.
On all graphs with $m > 10^5$, the parallel algorithm is faster than the sequential algorithms and the observed speed-up increases with the size of the graph.

\subsubsection{Scaling}
\begin{figure}[!t]
 \centering
 \includegraphics[width=0.8\textwidth]{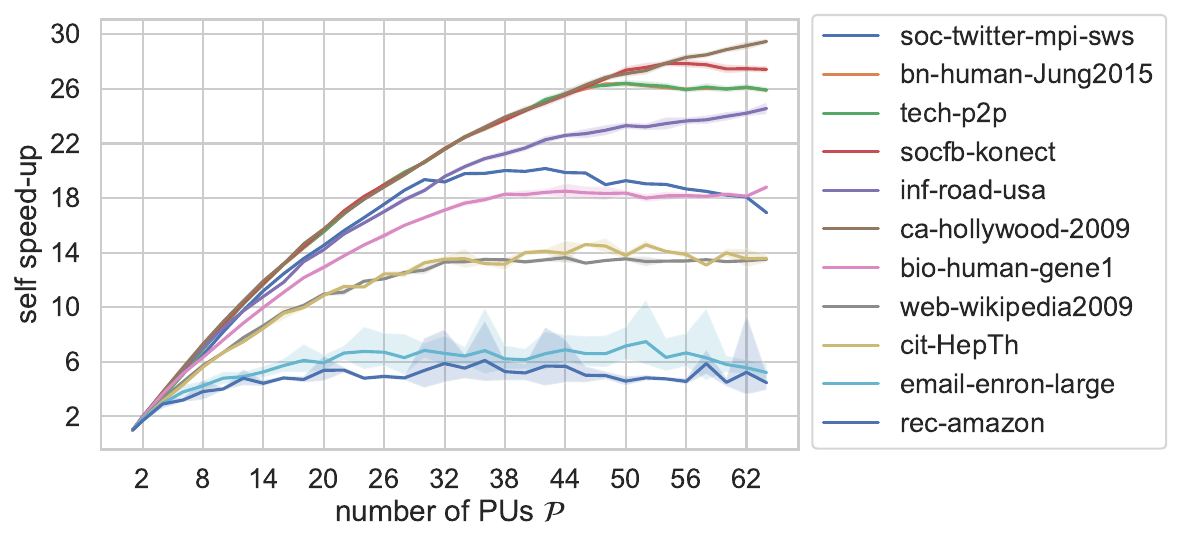}
 \caption{Strong scaling of \trueparglobes on a sample of graphs from \textsc{NetRep} for $1 \leq \nproc \leq~64$. The line colors indicate the graph and are sorted by graph size.}
 \label{fig:scaling-p-netrep-sample}
\end{figure}

We first report the self speed-up of \trueparglobes on the sample of graphs from \textsc{NetRep} in dependence of $\nproc$ (\cref{fig:scaling-p-netrep-sample}).
For larger graphs, the maximum speed-up ranges between $20$ and $30$ using $32$ to $64$ PUs.
On the two smallest graphs, the work is likely too small to be efficiently parallelized (e.g. compare \cref{table:rt-netrep-sample}).
An outlier is the largest graph (soc-twitter-mpi-sws); a likely explanation for the slightly low speed-up on this graph is that the highly skewed degree sequence increases the number of target dependencies and the synchronization overhead.

\begin{figure}[!t]
 \centering
 \includegraphics[width=0.7\textwidth]{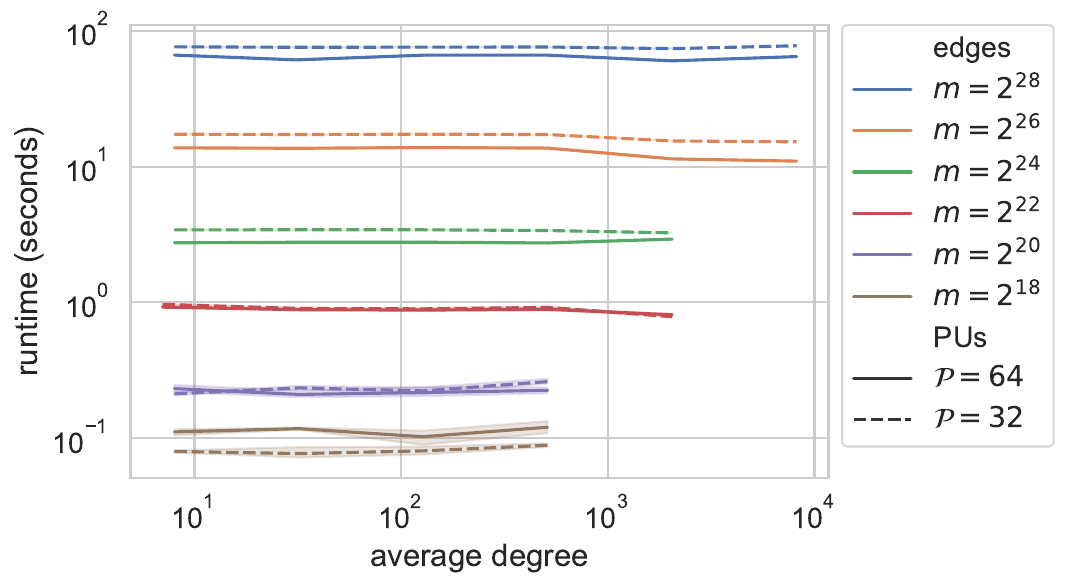}
 \caption{Runtime of \trueparglobes on graphs from \textsc{SynGnp} where $m~\in~\{2^{18}, 2^{20}, 2^{22}, 2^{24}, 2^{26}\}$ in dependence of the average degree $\overline{d}~=~2 m / n$. The color indicates the number of edges and the line style the number of PUs.}
 \label{fig:scaling-avg-d}
\end{figure}
 
\begin{figure}[!t]
 \centering
 \includegraphics[width=0.7\textwidth]{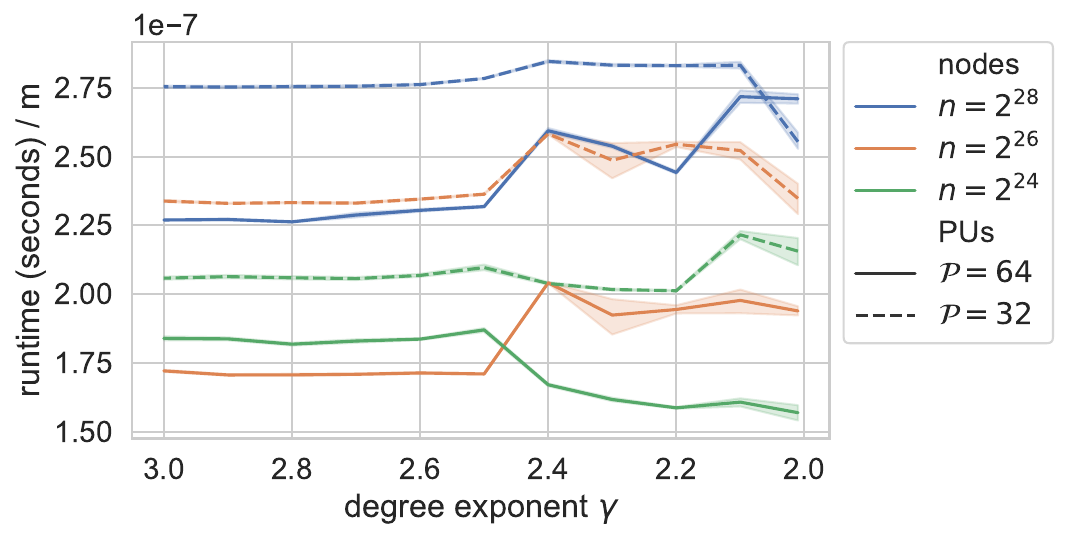}
 \caption{Runtime per edge of \trueparglobes on graphs from \textsc{SynPld} where $n \in \{2^{24}, 2^{26}, 2^{28}\}$ in dependence of the degree exponent $\gamma$. The color indicates the number of nodes and the line style the number of PUs.}
 \label{fig:scaling-gamma}
\end{figure}

\begin{figure}
 \centering
 \includegraphics[width=0.7\textwidth]{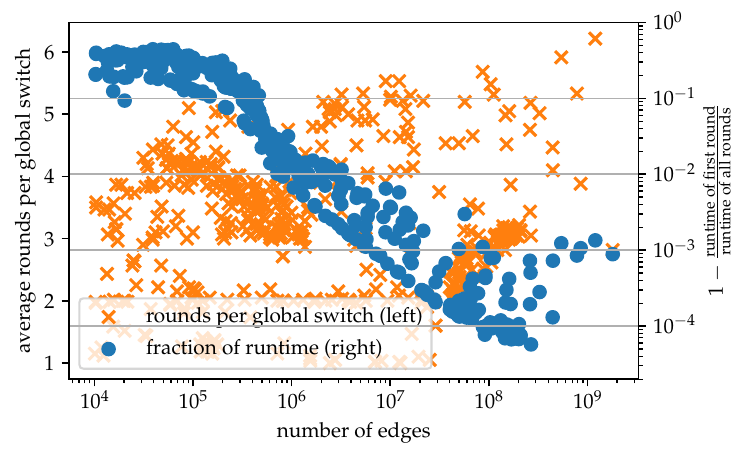}
	\caption{
		Rounds per global switch performed by \trueparglobes on \textsc{NetRep}.
		Each orange cross corresponds the average number of rounds recorded for a graph;
		the blue dots indicate the fractional runtime of all rounds excluding the first one.
	}
	\label{fig:real-rounds}
\end{figure}

To measure the influence of the graph properties on the runtime, we consider synthetic graphs.
We first consider graphs from \textsc{SynGnp} for various $n$ and $p$, and plot the runtime as a function of the average degree $\overline{d} = 2 m / n$ in \cref{fig:scaling-avg-d}.
The edge probability $p$ seems to have no significant effect on the runtime, even in the case where the average degree approaches the possible maximum $n - 1$ (bottom-right in the plot).
This is a consequence of \cref{thm:std-g-es-scales}.
As $G(n,p)$ graphs are sufficiently close to regular with high probability, the number of rounds required to perform a global switch is constant regardless of $p$.

Next, we consider power-law graphs from \textsc{SynPld} with degree exponent $3 \geq \gamma \geq~2.01$ to evaluate the influence of the degree distribution's skewness.
Note that increasing $\gamma$ increases the number of edges even when fixing $n$, therefore we normalize the runtime by dividing by the number of edges.
We report the runtime per edge as a function of $\gamma$ in \cref{fig:scaling-gamma}.
We observe an effect both in $n$ and $\gamma$.
For $n \geq 2^{26}$, the runtime on 64 PUs increases slightly as $\gamma$ approaches~$2$.
This matches the analysis given for \cref{thm:std-g-es-scales2}: for graphs with a highly skewed degree sequence, most edge switches will attempt to create the same few edges, causing many target dependencies and more synchronization overhead.

\subsubsection{Rounds per Global Switch}

Recall that \trueparglobes may delay edge switches to resolve target dependencies.
It does so by executing several rounds.
To study the performance impact, we execute 20 global switches per graph of \textsc{NetRep} using $\nproc=32$~PUs and record the number of rounds per global switch, and the time accumulated on all rounds excluding the first one.
As reported in \cref{fig:real-rounds}, the average number of rounds is low with a mean of $2.2$, and the maximum number of rounds we observed was $8$.
For all networks with more than 4M edges the first round also accounts for more than $99\%$ of the runtime.
This suggests that even in a case where more rounds are required, the performance impact of the following rounds is negligible for sufficiently large graphs.

\section{Conclusions}
\label{sec:conclusions}
We propose \gesmc as an \esmc variant that exhibits more parallelism due to the absence of source dependencies.
In addition, we propose \truepares and \trueparglobes as exact parallelizations of \esmc and \gesmc.
These algorithms avoid possible deviations from the intended random process by accounting for the complex dependencies between edge switches.
To the best of our knowledge, they are the first exact parallelizations of switch Markov Chains for the uniform sampling of simple undirected graphs.

An autocorrelation analysis suggests that \gesmc typically requires fewer steps than standard \esmc to randomize a graph.
On $\nproc=32$ PUs, our parallel algorithm executes $10-12$ times faster than our sequential \gesmc implementation, and $50-100$ faster than existing \esmc implementations.
We investigate the number of rounds needed for \trueparglobes to perform a global switch and find that very few rounds are required in practice.
The experiments on the influence of graph properties match the theoretical predictions.
For regular graphs, the performance is not affected by the density of the graph.
For power-law graphs with very small degree exponents $\gamma < 2.2$, there is a slight slowdown, due to the increased number of target dependencies.
However, in our experiment on over $600$ real graphs, this occurs for only very few outliers.

We expect that dedicated base cases for small graphs can further reduce the overhead due to the synchronization and concurrent data structures and thereby improve the scaling on such graphs.
We are also still interested in analyzing if \gesmc is rapidly mixing for any class of undirected graphs.
While we expect lower speed-ups than for \trueparglobes, an alternative could be to investigate the scalability of \truepares since this algorithm inherits the rapid mixing property of \esmc.

\section*{Acknowledgments}
Extensive calculations on the Goethe-HLR high-performance computer of the Goethe University Frankfurt were conducted for this research.
The authors would like to acknowledge the CSC team for their support.

\bibliographystyle{elsarticle-num}

\bibliography{par-es}

\begin{thebibliography}{10}

\bibitem{DBLP:journals/im/BoldiV14}
Paolo Boldi and Sebastiano Vigna.
\newblock Axioms for centrality.
\newblock {\em Internet Math.}, 2014.

\bibitem{barabasi2016network}
Albert-L{\'a}szl{\'o} Barab{\'a}si and M{\'a}rton P{\'o}sfai.
\newblock {\em Network science}.
\newblock 2016.

\bibitem{Cobb2003}
G.~W. Cobb and Y.-P. Chen.
\newblock An application of {M}arkov {C}hain {M}onte {C}arlo to community
  ecology.
\newblock {\em American Math. Monthly}, 110, 2003.

\bibitem{Itzkovitz2003}
S.~Itzkovitz, R.~Milo, N.~Kashtan, G.~Ziv, and U.~Alon.
\newblock Subgraphs in random networks.
\newblock {\em Phys. Rev. E}, 68, 2003.

\bibitem{DBLP:conf/asunam/SchlauchZTA15}
Wolfgang~Eugen Schlauch and Katharina~Anna Zweig.
\newblock Influence of the null-model on motif detection.
\newblock In {\em {ASONAM}}, 2015.

\bibitem{CarstensPhd}
C.~J. Carstens.
\newblock {\em {Topology of Complex Networks: Models and Analysis}}.
\newblock PhD thesis, RMIT University, 2016.

\bibitem{DBLP:journals/corr/abs-2003-00736}
Manuel Penschuck, Ulrik Brandes, Michael Hamann, Sebastian Lamm, Ulrich Meyer,
  Ilya Safro, Peter Sanders, and Christian Schulz.
\newblock Recent advances in scalable network generation.
\newblock {\em CoRR}, abs/2003.00736, 2020.

\bibitem{Havel1955}
V\'{a}clav Havel.
\newblock Pozn\'{a}mka o existenci kone\v{c}n\'{y}ch graf\r{u}.
\newblock {\em \v{C}asopis pro p\v{e}stov\'{a}n\'{i} matematiky}, 080, 1955.

\bibitem{doi:10.1137/0110037}
Seifollah~L. Hakimi.
\newblock On realizability of a set of integers as degrees of the vertices of a
  linear graph. i.
\newblock {\em J. SIAM}, 10, 1962.

\bibitem{DBLP:journals/im/BlitzsteinD11}
Joseph~K. Blitzstein and Persi Diaconis.
\newblock A sequential importance sampling algorithm for generating random
  graphs with prescribed degrees.
\newblock {\em Internet Math.}, 6, 2011.

\bibitem{DBLP:conf/bigdataconf/BhuiyanKM17}
Md~Hasanuzzaman Bhuiyan, Maleq Khan, and Madhav Marathe.
\newblock A parallel algorithm for generating a random graph with a prescribed
  degree sequence.
\newblock In {\em {IEEE} BigData}, 2017.

\bibitem{chung2002connected}
F.~Chung and L.~Lu.
\newblock Connected components in random graphs with given expected degree
  sequences.
\newblock {\em A. of Comb.}, 6, 2002.

\bibitem{DBLP:journals/datamine/MorenoPN18}
Sebasti{\'{a}}n Moreno, Joseph J.~Pfeiffer III, and Jennifer Neville.
\newblock Scalable and exact sampling method for probabilistic generative graph
  models.
\newblock {\em Data Min. Knowl. Discov.}, 32, 2018.

\bibitem{DBLP:journals/jct/BenderC78}
E.~A. Bender and E.~R. Canfield.
\newblock The asymptotic number of labeled graphs with given degree sequences.
\newblock {\em J. of Comb. Theo. {A}}, 1978.

\bibitem{bekessy1972asymptotic}
A.~B{\'e}k{\'e}ssy, P.~B{\'e}k{\'e}ssy, and J.~Koml{\'o}s.
\newblock Asymptotic enumeration of regular matrices.
\newblock {\em Stud. Sci. Math. Hungar.}, 7, 1972.

\bibitem{DBLP:books/ox/Newman10}
M.~E.~J. Newman.
\newblock {\em Networks: An Introduction}.
\newblock 2010.

\bibitem{bollobas1985random}
B.~Bollob{\'{a}}s.
\newblock {\em Random graphs}.
\newblock 1985.

\bibitem{DBLP:journals/ejc/Bollobas80}
B.~Bollob\'{a}s.
\newblock A probabilistic proof of an asymptotic formula for the number of
  labelled regular graphs.
\newblock {\em Eur. J. Comb.}, 1, 1980.

\bibitem{DBLP:journals/jal/McKayW90}
B.~D. McKay and N.~C. Wormald.
\newblock Uniform generation of random regular graphs of moderate degree.
\newblock {\em J. Algorithms}, 11, 1990.

\bibitem{DBLP:conf/focs/GaoW15}
P.~Gao and N.~C. Wormald.
\newblock Uniform generation of random regular graphs.
\newblock In {\em {FOCS}}, 2015.

\bibitem{DBLP:conf/focs/ArmanGW19}
A.~Arman, P.~Gao, and N.~C. Wormald.
\newblock Fast uniform generation of random graphs with given degree sequences.
\newblock In {\em {FOCS}}, 2019.

\bibitem{DBLP:journals/tcs/JerrumS90}
M.~Jerrum and A.~Sinclair.
\newblock Fast uniform generation of regular graphs.
\newblock {\em TCS}, 73, 1990.

\bibitem{DBLP:journals/cpc/CooperDG07}
C.~Cooper, M.~E. Dyer, and C.~S. Greenhill.
\newblock Sampling regular graphs and a peer-to-peer network.
\newblock {\em Comb. Probab. Comput.}, 2007.

\bibitem{DBLP:conf/alenex/GkantsidisMMZ03}
Christos Gkantsidis, Milena Mihail, and Ellen~W. Zegura.
\newblock The {M}arkov {C}hain simulation method for generating connected power
  law random graphs.
\newblock In {\em {ALENEX}}, 2003.

\bibitem{DBLP:conf/soda/Greenhill15}
C.~S. Greenhill.
\newblock The switch {M}arkov {C}hain for sampling irregular graphs.
\newblock In {\em {SODA}}, 2015.

\bibitem{DBLP:journals/rsa/KannanTV99}
R.~Kannan, P.~Tetali, and S.~S. Vempala.
\newblock Simple {M}arkov {C}hain algorithms for generating bipartite graphs
  and tournaments.
\newblock {\em RSA}, 1999.

\bibitem{DBLP:conf/sigcomm/MahadevanKFV06}
P.~Mahadevan, D.~V. Krioukov, K.~R. Fall, and V.~Vahdat.
\newblock Systematic topology analysis and generation using degree
  correlations.
\newblock In {\em {SIGCOMM}}, 2006.

\bibitem{DBLP:conf/alenex/StantonP11}
I.~Stanton and A.~Pinar.
\newblock Sampling graphs with a prescribed joint degree distribution using
  {M}arkov {C}hains.
\newblock In {\em {ALENEX}}, 2011.

\bibitem{strona2014fast}
G.~Strona, D.~Nappo, F.~Boccacci, S.~Fattorini, and J.~San-Miguel-Ayanz.
\newblock A fast and unbiased procedure to randomize ecological binary matrices
  with fixed row and column totals.
\newblock {\em Nature comm.}, 2014.

\bibitem{verhelst2008efficient}
N.~D. Verhelst.
\newblock An efficient {MCMC} algorithm to sample binary matrices with fixed
  marginals.
\newblock {\em Psychometrika}, 73, 2008.

\bibitem{DBLP:journals/compnet/VigerL16}
F.~Viger and M.~Latapy.
\newblock Efficient and simple generation of random simple connected graphs
  with prescribed degree sequence.
\newblock {\em J. Complex Networks}, 4, 2016.

\bibitem{10.1371/journal.pone.0131300}
Péter~L. Erdős, Sándor~Z. Kiss, István Miklós, and Lajos Soukup.
\newblock Approximate counting of graphical realizations.
\newblock {\em {PLOS ONE}}, 10, 2015.

\bibitem{DBLP:journals/tcs/GreenhillS18}
Catherine~S. Greenhill and Matteo Sfragara.
\newblock The switch {M}arkov {C}hain for sampling irregular graphs and
  digraphs.
\newblock {\em TCS}, 719, 2018.

\bibitem{ERDOS2022103421}
Péter~L. Erdős, Catherine Greenhill, Tamás~Róbert Mezei, István Miklós,
  Dániel Soltész, and Lajos Soukup.
\newblock The mixing time of switch markov chains: A unified approach.
\newblock {\em Eur. J. Comb.}, 99, 2022.

\bibitem{DBLP:journals/rsa/AmanatidisK20}
Georgios Amanatidis and Pieter Kleer.
\newblock Rapid mixing of the switch {M}arkov {C}hain for strongly stable
  degree sequences.
\newblock {\em RSA}, 57, 2020.

\bibitem{DBLP:journals/dam/GaoG21}
Pu~Gao and Catherine~S. Greenhill.
\newblock Mixing time of the switch {M}arkov {C}hain and stable degree
  sequences.
\newblock {\em Discret. Appl. Math.}, 291, 2021.

\bibitem{DBLP:journals/ejc/ErdosGMMSS22}
P{\'{e}}ter~L. Erd{\"{o}}s, Catherine~S. Greenhill, Tam{\'{a}}s~R{\'{o}}bert
  Mezei, Istv{\'{a}}n Mikl{\'{o}}s, Daniel Solt{\'{e}}sz, and Lajos Soukup.
\newblock The mixing time of switch markov chains: {A} unified approach.
\newblock {\em Eur. J. Comb.}, 99:103421, 2022.

\bibitem{Milo2003}
R.~Milo, N.~Kashtan, S.~Itzkovitz, M.~E.~J. Newman, and U.~Alon.
\newblock On the uniform generation of random graphs with prescribed degree
  sequences.
\newblock {\em CoRR}, abs/cond-mat/0312028, 2003.

\bibitem{DBLP:conf/waw/RayPS12}
Jaideep Ray, Ali Pinar, and C.~Seshadhri.
\newblock Are we there yet? when to stop a {M}arkov {C}hain while generating
  random graphs.
\newblock In {\em {WAW}}, volume 7323 of {\em {LNCS}}, 2012.

\bibitem{DBLP:journals/jpdc/BhuiyanKCM17}
Md~Hasanuzzaman Bhuiyan, Maleq Khan, Jiangzhuo Chen, and Madhav~V. Marathe.
\newblock Parallel algorithms for switching edges in heterogeneous graphs.
\newblock {\em J. Parallel Distributed Comput.}, 104, 2017.

\bibitem{9820710}
D.~Allendorf, U.~Meyer, M.~Penschuck, and H.~Tran.
\newblock Parallel global edge switching for the uniform sampling of simple
  graphs with prescribed degrees.
\newblock In {\em 2022 IEEE International Parallel and Distributed Processing
  Symposium (IPDPS)}, pages 269--279, Los Alamitos, CA, USA, jun 2022. IEEE
  Computer Society.

\bibitem{DBLP:conf/esa/CarstensH0PTW18}
{C. J. Carstens}, M.~Hamann, U.~Meyer, M.~Penschuck, H.~Tran, and D.~Wagner.
\newblock Parallel and {I/O}-efficient randomisation of massive networks using
  {G}lobal {C}urveball trades.
\newblock {\em {ESA}}, 2018.

\bibitem{pandey2020c}
Santosh Pandey, Lingda Li, Adolfy Hoisie, Xiaoye~S Li, and Hang Liu.
\newblock C-saw: A framework for graph sampling and random walk on gpus.
\newblock In {\em SC20: International Conference for High Performance
  Computing, Networking, Storage and Analysis}, pages 1--15. IEEE, 2020.

\bibitem{DBLP:journals/jea/HamannMPTW18}
Michael Hamann, Ulrich Meyer, Manuel Penschuck, Hung Tran, and Dorothea Wagner.
\newblock {I/O}-efficient generation of massive graphs following the \emph{LFR}
  benchmark.
\newblock {\em {ACM} J. Exp. Algorithmics}, 23, 2018.

\bibitem{DBLP:journals/corr/abs-2108-11613}
Artur Czumaj and Andrzej Lingas.
\newblock On truly parallel time in population protocols.
\newblock {\em CoRR}, abs/2108.11613, 2021.

\bibitem{DBLP:journals/corr/CarstensBS16}
Corrie~Jacobien Carstens, Annabell Berger, and Giovanni Strona.
\newblock {C}urveball: a new generation of sampling algorithms for graphs with
  fixed degree sequence.
\newblock {\em CoRR}, abs/1609.05137, 2016.

\bibitem{DBLP:books/daglib/0012859}
Michael Mitzenmacher and Eli Upfal.
\newblock {\em Probability and Computing: Randomized Algorithms and
  Probabilistic Analysis}.
\newblock 2005.

\bibitem{rao1996}
A.~Ramachandra Rao, Rabindranath Jana, and Suraj Bandyopadhyay.
\newblock A {M}arkov {C}hain {M}onte {C}arlo method for generating random $(0,
  1)$-matrices with given marginals.
\newblock {\em Sankhyā: The Indian J. of Statistics A}, 58, 1996.

\bibitem{DBLP:conf/approx/CarstensK18}
Corrie~Jacobien Carstens and Pieter Kleer.
\newblock Speeding up switch {M}arkov {C}hains for sampling bipartite graphs
  with given degree sequence.
\newblock In {\em {APPROX/RANDOM}}, volume 116 of {\em LIPIcs}, 2018.

\bibitem{DBLP:conf/bcc/Greenhill21}
Catherine~S. Greenhill.
\newblock Generating graphs randomly.
\newblock In Konrad~K. Dabrowski, Maximilien Gadouleau, Nicholas Georgiou,
  Matthew Johnson, George~B. Mertzios, and Dani{\"{e}}l Paulusma, editors, {\em
  Surveys in Combinatorics, 2021: Invited lectures from the 28th British
  Combinatorial Conference, Durham, UK, July 5-9, 2021}, pages 133--186.
  Cambridge University Press, 2021.

\bibitem{DBLP:journals/netsci/StaudtSM16}
C.~L. Staudt, A.~Sazonovs, and H.~Meyerhenke.
\newblock Networkit: {A} tool suite for large-scale complex network analysis.
\newblock {\em Netw. Sci.}, 4, 2016.

\bibitem{SciPyProceedings_11}
Aric~A. Hagberg, Daniel~A. Schult, and Pieter~J. Swart.
\newblock Exploring network structure, dynamics, and function using {NetworkX}.
\newblock In {\em {SciPy}}, 2008.

\bibitem{Lancichinetti2009}
A.~Lancichinetti and S.~Fortunato.
\newblock Benchmarks for testing community detection algorithms on directed and
  weighted graphs with overlapping communities.
\newblock {\em Phys. Rev. E}, 80, 2009.

\bibitem{DBLP:journals/corr/abs-cs-0502085}
Fabien Viger and Matthieu Latapy.
\newblock Fast generation of random connected graphs with prescribed degrees.
\newblock {\em CoRR}, abs/cs/0502085, 2005.

\bibitem{intel-manual-19}
Intel Corporation.
\newblock {\em Intel\textregistered 64 and {IA-32} Architectures --- Software
  Developer’s Manual --- Vol. 2}, 2019.

\bibitem{DBLP:journals/topc/MaierSD19}
Tobias Maier, Peter Sanders, and Roman Dementiev.
\newblock Concurrent hash tables: Fast and general(?)!
\newblock {\em {ACM} Trans. Parallel Comput.}, 5(4), 2019.

\bibitem{DBLP:journals/tomacs/MatsumotoN98}
Makoto Matsumoto and Takuji Nishimura.
\newblock Mersenne twister: {A} 623-dimensionally equidistributed uniform
  pseudo-random number generator.
\newblock {\em {ACM} Trans. Model. Comput. Simul.}, 8(1), 1998.

\bibitem{DBLP:journals/tomacs/Lemire19}
D.~Lemire.
\newblock Fast random integer generation in an interval.
\newblock {\em {ACM} Trans. Model. Comput. Simul.}, 29, 2019.

\bibitem{DBLP:journals/ipl/Sanders98}
P.~Sanders.
\newblock Random permutations on distributed, external and hierarchical memory.
\newblock {\em Inf. Process. Lett.}, 67, 1998.

\bibitem{Gilbert59}
E.~N. Gilbert.
\newblock {Random Graphs}.
\newblock {\em Ann. Math. Stat.}, 30(4), 1959.

\bibitem{DBLP:conf/soda/GaoW18}
P.~Gao and N.~C. Wormald.
\newblock Uniform generation of random graphs with power-law degree sequences.
\newblock In {\em {SODA}}, 2018.

\bibitem{DBLP:conf/aaai/RossiA15}
R.~A. Rossi and N.~K. Ahmed.
\newblock The network data repository with interactive graph analytics and
  visualization.
\newblock In {\em {AAAI}}, 2015.

\bibitem{DBLP:journals/jea/StantonP11}
Isabelle Stanton and Ali Pinar.
\newblock Constructing and sampling graphs with a prescribed joint degree
  distribution.
\newblock {\em {ACM} J. Exp. Algorithmics}, 17.

\bibitem{DBLP:journals/compnet/RayPS15}
Jaideep Ray, Ali Pinar, and C.~Seshadhri.
\newblock A stopping criterion for {M}arkov {C}hains when generating
  independent random graphs.
\newblock {\em J. Complex Networks}, 3, 2015.

\bibitem{berger2018smaller}
Annabell Berger and Corrie~Jacobien Carstens.
\newblock Smaller universes for uniform sampling of 0,1-matrices with fixed row
  and column sums.
\newblock {\em CoRR}, abs/1803.02624, 2018.

\bibitem{bishop2007discrete}
Yvonne~M Bishop, Stephen~E Fienberg, and Paul~W Holland.
\newblock {\em Discrete multivar. analysis: theo. and practice}.
\newblock 2007.

\end{thebibliography}

\end{document}